\documentclass[pdflatex,sn-mathphys-num]{sn-jnl}

\usepackage{graphicx}%
\usepackage{array}
\usepackage{multirow}%
\usepackage{amsmath,amssymb,amsfonts}%
\usepackage{xcolor}%
\usepackage{orcidlink}

\usepackage{tikz}
\newcommand\copyrighttext{%
  \scriptsize This version of the article has been accepted for publication, after peer review, but is not the Version of Record and does not reflect post-acceptance improvements, or any corrections. The Version of Record is published in the \textit{International Journal of Artificial Intelligence in Education (IJAIED)}, and is available online at \url{https://doi.org/10.1007/s40593-025-00478-6}.
  \smallskip
  
  \textcopyright\ 2025. Please cite this article as follows: E. D. López Zapata, C. Tang, V. Švábenský, F. Okubo, A. Shimada: \textit{LECTOR: Summarizing E-book Reading Content for Personalized Student Support}. In International Journal of Artificial Intelligence in Education (IJAIED), Springer Nature, 2025. DOI: \href{https://doi.org/10.1007/s40593-025-00478-6}{10.1007/s40593-025-00478-6}.}
\newcommand\copyrightnotice{%
\begin{tikzpicture}[remember picture,overlay]
\node[anchor=north,yshift=-24pt] at (current page.north) {\fbox{\parbox{\dimexpr\textwidth-\fboxsep-\fboxrule\relax}{\copyrighttext}}};
\end{tikzpicture}%
}

\begin{document}

\title[Article Title]{LECTOR: Summarizing E-book Reading Content for Personalized Student Support}


\author*[1]{\fnm{Erwin} \sur{López} \orcidlink{0000-0003-3793-9524}}\email{edlopez96s6@gmail.com}

\author[2]{\fnm{Cheng} \sur{Tang} \orcidlink{0000-0002-8148-1509}}\email{	
tang@ait.kyushu-u.ac.jp}

\author[2]{\fnm{Valdemar} \sur{Švábenský} \orcidlink{0000-0001-8546-280X}}\email{valdemar@kyudai.jp}

\author[2]{\fnm{Fumiya} \sur{Okubo} \orcidlink{0000-0002-0077-9072}}\email{fokubo@ait.kyushu-u.ac.jp}

\author[2]{\fnm{Atsushi} \sur{Shimada} \orcidlink{0000-0002-3635-9336}}\email{atsushi@ait.kyushu-u.ac.jp}

\affil*[1]{\orgdiv{Graduate School of Information Science and Electrical Engineering}, \orgname{Kyushu University}, \orgaddress{\street{744 Motooka, Nishi-ku}, \city{Fukuoka City}, \postcode{819-0395}, \state{Fukuoka}, \country{Japan}}}

\affil[2]{\orgdiv{Faculty of Information Science and Electrical Engineering}, \orgname{Kyushu University}, \orgaddress{\street{744 Motooka, Nishi-ku}, \city{Fukuoka City}, \postcode{819-0395}, \state{Fukuoka}, \country{Japan}}}


\abstract{Educational e-book platforms provide valuable information to teachers and researchers through two main sources: reading activity data and reading content data. While reading activity data is commonly used to analyze learning strategies and predict low-performing students, reading content data is often overlooked in these analyses. To address this gap, this study proposes LECTOR (Lecture slides and Topic Relationships), a model that summarizes information from reading content in a format that can be easily integrated with reading activity data. Our first experiment compared LECTOR to representative Natural Language Processing (NLP) models in extracting key information from 2,255 lecture slides, showing an average improvement of 5\% in F1-score. These results were further validated through a human evaluation involving 28 students, which showed an average improvement of 21\% in F1-score over a model predominantly used in current educational tools. Our second experiment compared reading preferences extracted by LECTOR with traditional reading activity data in predicting low-performing students using 600,712 logs from 218 students. The results showed a tendency to improve the predictive performance by integrating LECTOR. Finally, we proposed examples showing the potential application of the reading preferences extracted by LECTOR in designing personalized interventions for students.

\copyrightnotice}

\keywords{e-book reading logs, reading behavior analysis, student performance prediction, multimodal integration, topic extraction, student support}



\maketitle

\section{Introduction}\label{sec1}

The adoption of digital textbooks has been steadily increasing in formal education \cite{cadd_broad1, cadd_broad2}, including universities and K-12 classrooms in countries such as Japan \cite{cadd_jpn,ogata_2017}, South Korea \cite{cadd_kor}, Egypt \cite{cadd_egp}, Canada \cite{cadd_cnd}, and the USA \cite{cadd_usa_1,cadd_usa_2}. This adoption responds to growing efforts to improve accessibility and explore the potential of AI-driven systems to further enhance students' learning experiences \cite{cadd_broad1,cadd_broad2,cadd_broad3}. This potential is made possible by the reading logs generated by e-readers, which capture fine-grained records of student engagement with learning materials \cite{freeman_16,ogata_2017,flanagan_2017}. These logs can be leveraged to infer reading behavior and predict academic performance, ultimately informing educational interventions. However, the extent to which they can meaningfully support instructors in guiding students remains an open challenge \cite{cadd_broad2}.

Existing research on reading behavior analysis primarily focuses on clustering students based on behavioral features, such as reading time and navigation patterns, to identify distinct reading strategies \cite{akcapinar_chen_2020,akcapinar_hasnine_2020,yin_2019}. In contrast, student performance prediction models use supervised Machine Learning (ML) on larger feature sets to estimate the likelihood of students receiving low final grades (at-risk students) \cite{murata_2023,chen_2021,akcapinar_2019}. While both approaches provide valuable insights, they primarily rely on activity frequency data (e.g., under- or over-engagement) without considering the semantic content of the materials being read \cite{wang_2022, lopez_2023}. For example, at-risk prediction models can highlight students who need to engage more with lecture materials but do not specify which content requires review \cite{murata_2023,chen_2021,akcapinar_2019}.

\begin{figure}[h]
  \centering
  \includegraphics[height=2.3in, width=4.5in]{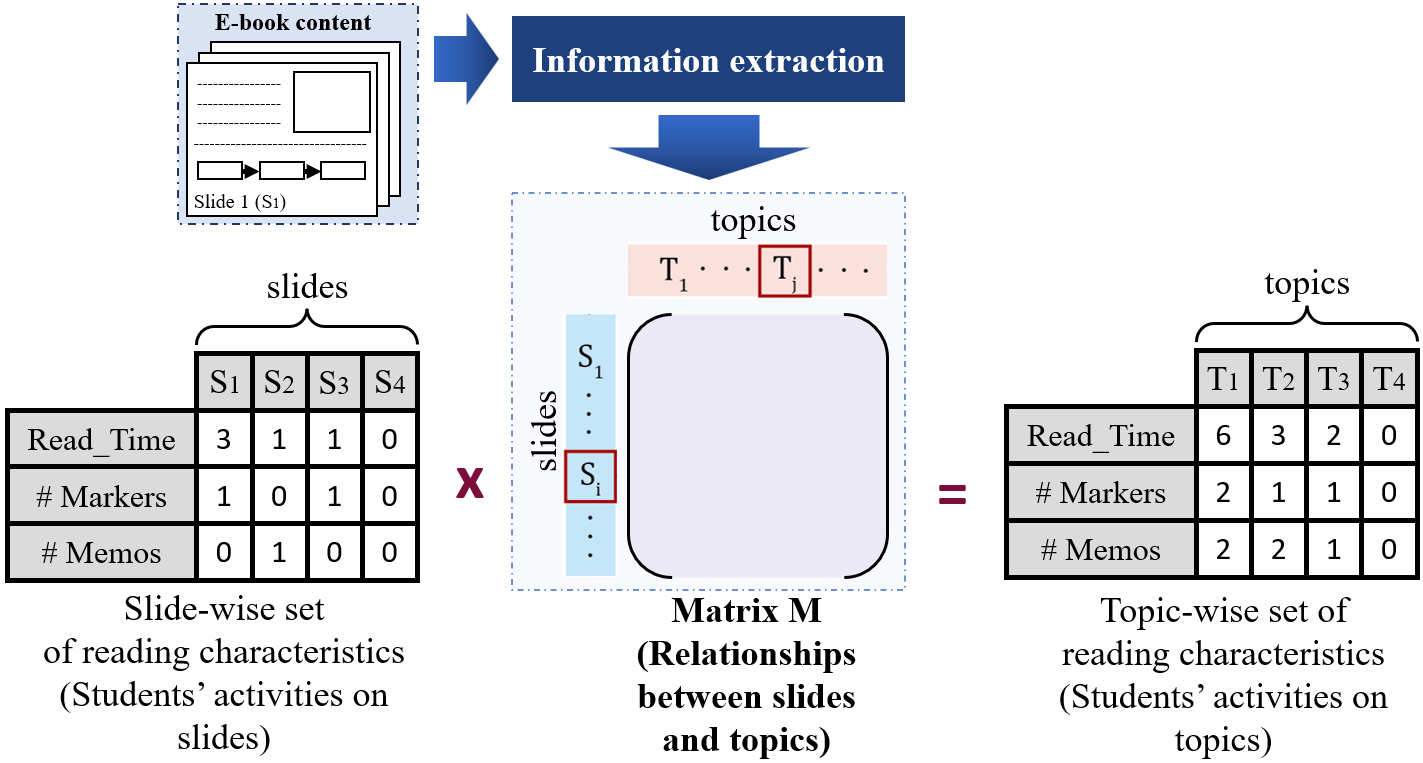}
  \caption{Our proposed multimodal integration multiplies the Slide-based reading characteristics by a Matrix M of Slide-Topic relationships to obtain Topic-based reading characteristics.}
  \label{fig:fig_1}
\end{figure}

A promising approach to address this limitation is to contextualize reading logs with the content of the corresponding materials, enabling a deeper understanding of how students interact with specific topics \cite{solar}. While this multimodal integration has not been formally defined, previous exploratory studies \cite{wang_2022, takii_2024,lopez_2023} provide a foundation for interpreting it as a matrix multiplication between traditional reading logs and a matrix M that encodes topic distributions across pages of the reading material (Figure \ref{fig:fig_1}). This interpretation reframes the challenge of limited educational intervention due to non-contextualized reading logs as an optimization problem, where the goal is to accurately estimate the distribution of key topics across educational materials.

This estimation is an open problem in Natural Language Processing (NLP), closely related to keyphrase extraction and information retrieval \cite{salton_1988, mihalcea_2004, bougouin_2013, rose_2010, bennani_2018, sun_2020, ding_luo_2021}. It also aligns with prior educational research on extracting textual features from digital reading materials to develop learning-support tools, such as slide summarization and recommendation systems \cite{shimada_2018, nakayama_2019, yang_2018, okubo_2020}. Despite advances in both areas, existing approaches remain limited. State-of-the-art NLP techniques excel at information retrieval but are not optimized for structured, domain-specific materials such as lecture slides \cite{bennani_2018, sun_2020, ding_luo_2021}. Conversely, educational tools are specifically designed for reading materials but rely on basic retrieval methods that lack semantic understanding \cite{shimada_2018, nakayama_2019, yang_2018, okubo_2020}.

In this context, we build upon recent advances in NLP to develop LECTOR (LECture slides and TOpic Relationships), a model designed to leverage the intrinsic structure of lecture slides for optimizing the retrieval of key information. LECTOR estimates a Matrix of Slides-Topics Relationships M (Figure \ref{fig:fig_1}), enabling the extraction of semantically contextualized reading activities. By incorporating topic-based features, this approach enhances the interpretability of student reading behaviors, improving the accuracy of at-risk student predictions and facilitating their translation into personalized educational interventions tailored to individual reading preferences.

In this study, we evaluate LECTOR through both quantitative and qualitative analyses from two perspectives. First, from an information retrieval perspective, we compare LECTOR’s performance against state-of-the-art NLP models and traditional retrieval methods commonly used in learning support tools. Second, from an AI in education perspective, we assess how semantically enriched, topic-based features improve the prediction and interpretation of at-risk student behaviors compared to conventional activity features.

Accordingly, our study addresses the following research questions (RQ):
\begin{itemize}
    \item \textbf{RQ1.} Does incorporating educational considerations into an information retrieval system improve its performance compared to state-of-the-art NLP methods and traditional learning support tools for e-book readers?
    \item \textbf{RQ2.} Do topic-based features extracted from contextualized reading activities improve the prediction performance and explainability of at-risk student behaviors compared to traditional features from reading logs?
\end{itemize}

The main contributions of LECTOR to the state of the art of e-book reader applications used in educational environments are as follows:
\begin{itemize}
    \item LECTOR extracts important information from lecture slides better than the model predominantly used in current educational systems, contributing to the development of future educational tools based on text information retrieval (e.g., summarization and recommendation systems).

    \item LECTOR can be used to extract students' topic preferences from traditional reading activity data. Incorporating this new insight into the analysis of predictions of at-risk students contributes to the design of personalized interventions tailored to individual students' reading preferences.
\end{itemize}
While LECTOR was first introduced by the previous pilot study \cite{lopez_2023}, the present study extends that work in the following ways:
\begin{itemize}
    \item Verification of generalization: The previous paper \cite{lopez_2023} based their evaluations only on data from a course in 2019. In this study, we considered data from the same course over five years (2019 to 2023).

    \item Human evaluation: The previous paper \cite{lopez_2023} used a technical evaluation to measure the performance of models in extracting important topics. In this study, we added a human evaluation with students from the 2023 course.

    \item Verification of the improvement in predicting at-risk students: The previous paper~\cite{lopez_2023} conducted an analysis that suggested that prediction models could be improved by integrating LECTOR. In this study, we verified these improvements. 
\end{itemize}

\section{Related Work}\label{sec2}

\subsection{Text Processing of E-book Lecture Slides}\label{sec2_1}
The estimation of our proposed Matrix of Slide and Topic relationships is based on the extraction of text features from e-book slides. Accordingly, we have reviewed previous educational developments that address the extraction of these features. As shown in this section, most of them used the TF-IDF \cite{salton_1988}, a classic NLP model. 

For example, the study \cite{shimada_2018} developed a slide summarization system that extracted text features using the TF-IDF model, giving each word a score based on its frequency. They then estimated the importance of each slide by summing the scores of the words it contained. Similarly, \cite{nakayama_2019} developed a system to recommend learning resources (websites) by implementing the TF-IDF model. For each slide, instead of summing the scores of its words, they extracted the words with the highest scores and recommended websites covering topics related to those words. 

In a different context, the study \cite{yang_2018} approached the problem of transferring learning footprints when updating e-book lecture materials. For each slide, they used the TF-IDF model to extract a vector with its corresponding words' frequency scores. They then estimated a quantitative relationship between the original and updated slides using the cosine similarity of their vector representations. To transfer the learning footprints, they matched the most related pair of slides. 

The work \cite{okubo_2020} applied the same approach to match test questions and e-book lecture slides for personalized slide recommendations. Their system identified incorrectly answered questions and used the cosine similarity between question and slide vectors to quantify their relationship. Although this work used the TF-IDF for generating the recommendations, they also implemented the Doc2Vec model (Mikolov et al., 2013) to represent the quizzes and slides as vectors. Doc2Vec is a more advanced NLP model that considers contextual information of the text in its vector representations (also called vector embeddings). Accordingly, this model allowed a better estimation of the questions-slides quantitative relationships.

In summary, previous work has demonstrated the ability of NLP models to extract key information from slides \cite{shimada_2018,nakayama_2019} and to quantify slide-slide \cite{yang_2018} and question-slide relationships \cite{okubo_2020}. Although the TF-IDF model has been predominantly used in these developments, the study \cite{okubo_2020} introduced the idea that more advanced NLP models may be more useful for these tasks. Accordingly, we implemented a more advanced NLP model (LECTOR) to extract key information from slides and estimate slide-topic relationships.
\subsection{Keyphrase Extraction from General Documents}\label{sec2_2}
The design of LECTOR is based on previous developments for the unsupervised extraction of keyphrases. A general approach for this extraction is to estimate a quantitative relationship (score) between a set of documents and their keyphrase candidates (all written phrases). By considering lecture slides as documents and topics as keyphrase candidates, this methodology allows us to obtain our proposed matrix M (estimated scores between slides and topics).

The literature on unsupervised keyphrase extraction can be categorized into three main approaches, each of which achieves better results than its predecessor. The first group consists of statistical models that define the score of a word in a document as a proportion of its frequency, correcting this value with additional statistics of the word. For example, the TF-IDF method \cite{salton_1988} applied a term inverse to the general frequency (set of all documents) to avoid considering common words as keywords. Accordingly, this group includes the text-processing techniques for e-book slides mentioned in the previous section.

The models from the second group implemented graph-based models to estimate the relevance of words. The core idea is to extract a graph where words are vertices and edges represent the word co-occurrence in documents. The most representative work in this group is TextRank \cite{mihalcea_2004}, which introduced the mentioned idea and applied a final ranking model derived from the PageRank algorithm \cite{page_1998}. TopicRank \cite{bougouin_2013} followed the same approach, but defined topics as clusters of words, using them as vertices of the graph. Finally, RAKE \cite{rose_2010} introduced a new ranking idea by calculating word statistics from the co-occurrence matrix and assigning them as the words' final scores.

The third group is integrated by works that represent document words as embedding vectors. The works in this group represent the state-of-the-art in the field, with models such as EmbedRank \cite{bennani_2018}, SIFRank \cite{sun_2020}, and AttentionRank \cite{ding_luo_2021}. EmbedRank introduced the idea of using sentence embedding vectors to estimate the relationships between documents and keyphrase candidates with cosine similarity. Their embedding vectors were calculated with the Doc2Vec model \cite{mikolov_2013}. SIFRank followed this idea but proposed to consider contextual information by using the pre-trained Language Model ELMo \cite{peters_2018}. In addition, they used the Smooth Inverse Frequency (SIF) model~\cite{arora_2017} to estimate sentence embeddings. The last work, AttentionRank, also used a pre-trained Language Model, BERT \cite{devlin_2019}, but refined previous methods by using the self and cross-attention mechanisms in the estimation of the sentence embeddings.

The models from the third group benefited from the Deep Learning developments in the field of NLP, especially from the attention mechanism proposed by \cite{vaswani_2017} and the design of attention-based Language Models such as BERT and ELMo. For this reason, similar to SIFRank \cite{sun_2020} and AttentionRank \cite{ding_luo_2021}, our proposed model LECTOR exploits the advantages of the attention mechanism. In addition, LECTOR implements a data contextualization based on the unique structure of slides, adapting the general approach of unsupervised keyphrase extraction to the domain of lecture slides.

\subsection{Integration of E-book Reading Activity Data and Reading Content Data}\label{sec2_3}
\begin{figure}[h]
  \centering
  \includegraphics[height=2.6in, width=4.5in]{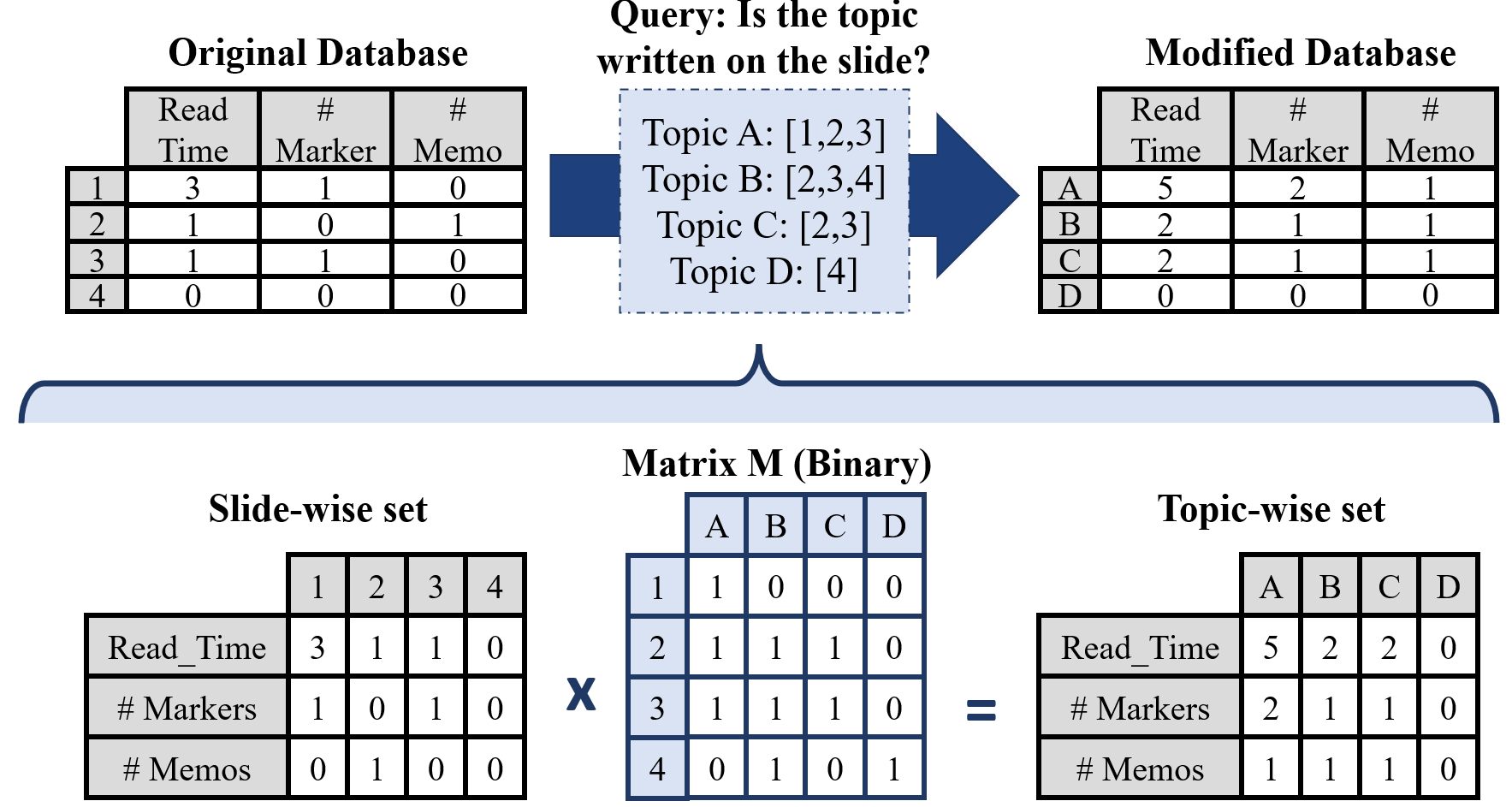}
  \caption{The methodology proposed by \cite{wang_2022} can be reformulated as a data transformation using a binary matrix of Slide-Topic relationships.}
  \label{fig:fig_2}
\end{figure}

Similar to our proposed approach, two recent studies have proposed integrating reading activity and content data. The study \cite{wang_2022} proposed this integration to analyze reader characteristics. Their method consisted of creating a list of all the slides where a topic appeared and querying this list to modify the original database of reading logs. As we show in Figure \ref{fig:fig_2}, this methodology can be reformulated as a data transformation using our proposed matrix M. Under this formulation, \cite{wang_2022} extracted text features by assigning a binary relationship of 1 if a topic appears in a given slide and 0 otherwise.

The study \cite{takii_2024} proposed this integration to create students' knowledge graphs (Open Knowledge and Learner Model). Their graph consisted of topics as graph nodes, topic-topic relationships as edges, and the number of reading logs as node sizes. The creation of this graph was based on the co-occurrence of topics in lecture slides, similar to TextRank \cite{mihalcea_2004}.

Since the main purpose of these two works was to discuss the potential of the reading activity and content data integration, they did not consider the technical performance of their NLP models. The work \cite{wang_2022} used a very simplified model (Binary score), and \cite{takii_2024} used a co-occurrence graph. In this study, we implement a more sophisticated NLP model (attention-based model) and validate the improvement over these previous models.

\section{Preliminaries}\label{sec3}
State-of-the-art works on unsupervised keyphrase extraction use attention-based models to extract important text features from the documents. Since our proposed model LECTOR follows the same approach, this section briefly introduces the attention mechanism.
\subsection{Self and Cross-Attention Mechanism}\label{sec3_1}
\begin{figure}[h]
  \centering
  \includegraphics[height=2.9in, width=2.1in]{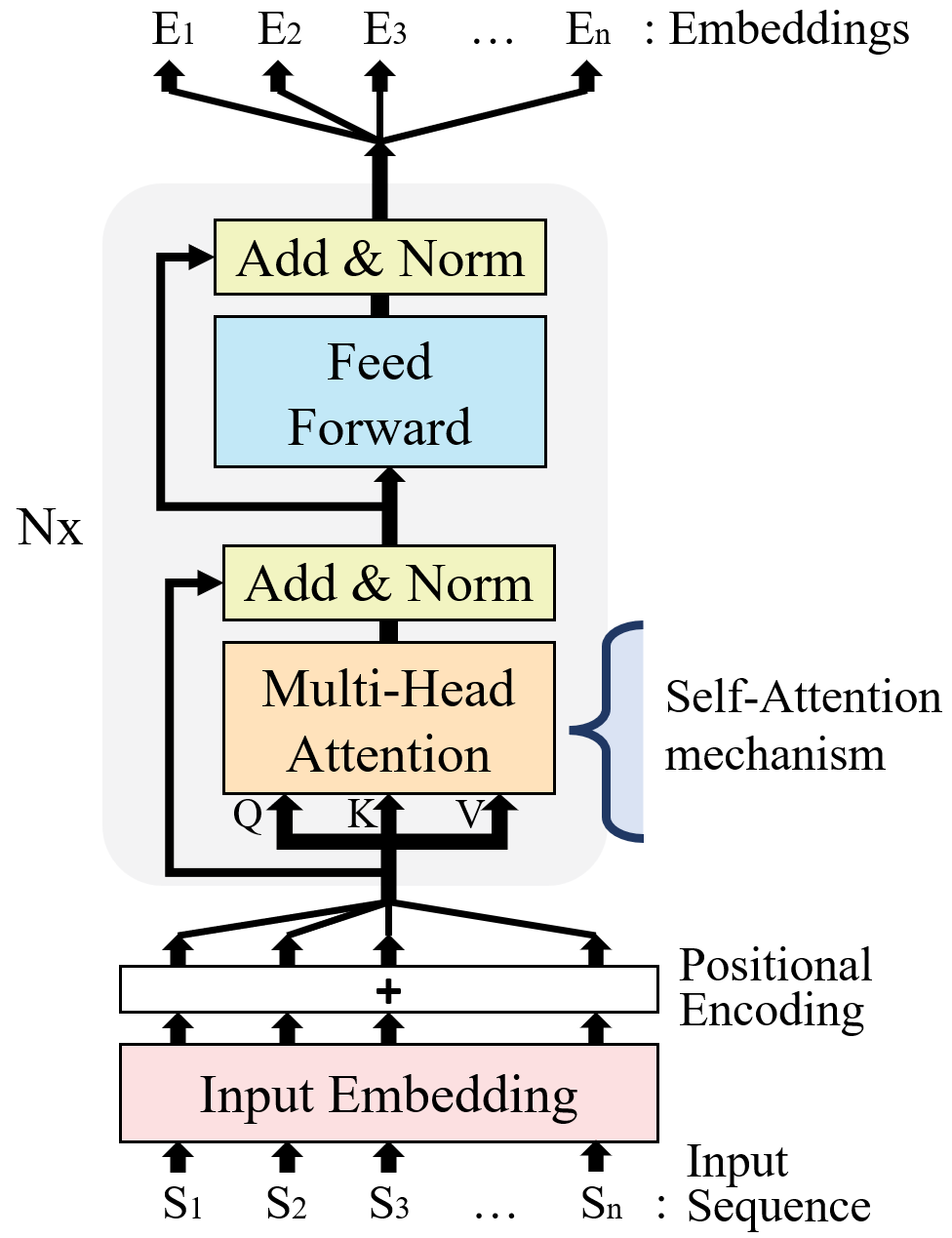}
  \caption{The architecture of the Transformer encoder, including the Input sequence, the Positional encoding, the Self-attention mechanism, and the generated set of Embeddings.}
  \label{fig:fig_3}
\end{figure}

As shown in Figure \ref{fig:fig_3}, a pre-trained Language Model has three main components: the Input Embedding, the Multi-Head Attention (MHA), and the Feed Forward layer. The first component linearly transforms the text input into a sequence of low-level embedding vectors. The MHA then extracts significant textual features from different parts of this sequence, and the third component encodes the extracted features into high-level embedding vectors.

The core of the MHA is the Self-attention mechanism and is defined as shown in Equation \ref{eq:self-attention_definition} \cite{vaswani_2017}. Here the Query (Q), Key (K), and Value (V) are linear transformations of the same input embedding.
\begin{equation}
\label{eq:self-attention_definition}
    Attention\left ( Q,K,V \right )=Softmax\left ( \frac{QK^{T}}{\sqrt{d_{k}}} \right )V
\end{equation}

\begin{figure}[h]
  \centering
  \includegraphics[height=2.8in, width=4.8in]{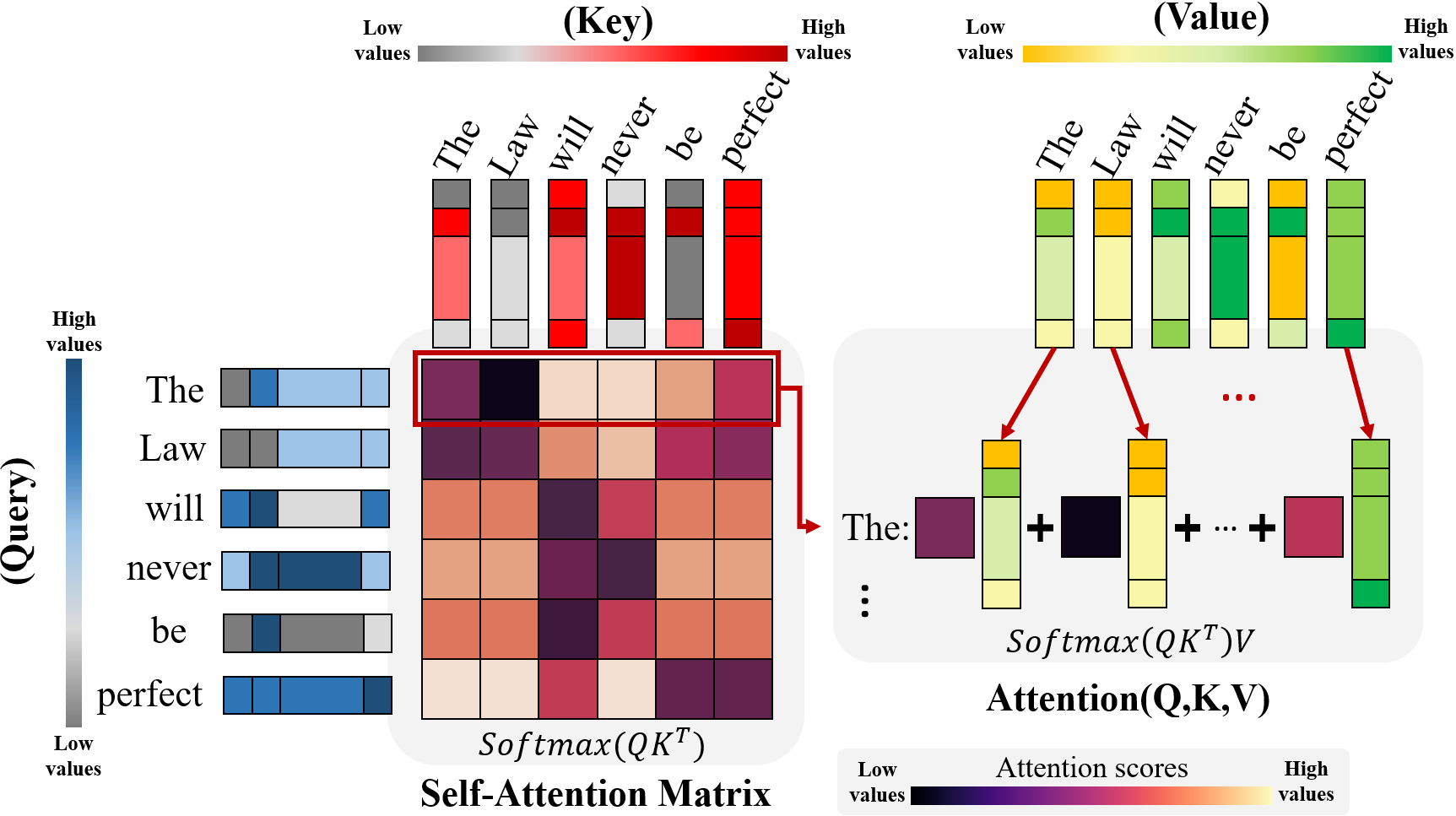}
  \caption{Example of the self-attention mechanism: The Self-Attention matrix quantifies the importance of a Key for a given Query, which is used to calculate a weighted average of the Value.}
  \label{fig:fig_4}
\end{figure}

In the first part of the equation, the Softmax function selects the “importance” (also called “attention”) of the Key for a given Query. Then, these importance scores are used to calculate a weighted sum of the Value. An example of this mechanism is shown in Figure \ref{fig:fig_4}. Here the Query, Key, and Value are generated from the sentence “The Law will never be perfect”. In the first part, we see the Self-attention Matrix, a square matrix that preserves the model's choices (attention values). Then, the final embeddings are calculated from the Value vectors with the highest attention.

The term “Self-attention” denotes that the Query and Key vectors represent the same input sentence. Accordingly, the Self-attention Matrix determines which parts of a sentence are significant in its own semantic context. However, in some cases, we need to consider two independent sentences in the attention mechanism. For example, in sentence translation, it is important to identify the key components of the original sentence in the translated language. In such cases, the Cross-attention mechanism is more effective.

\begin{figure}[h]
  \centering
  \includegraphics[height=1.9in, width=4.8in]{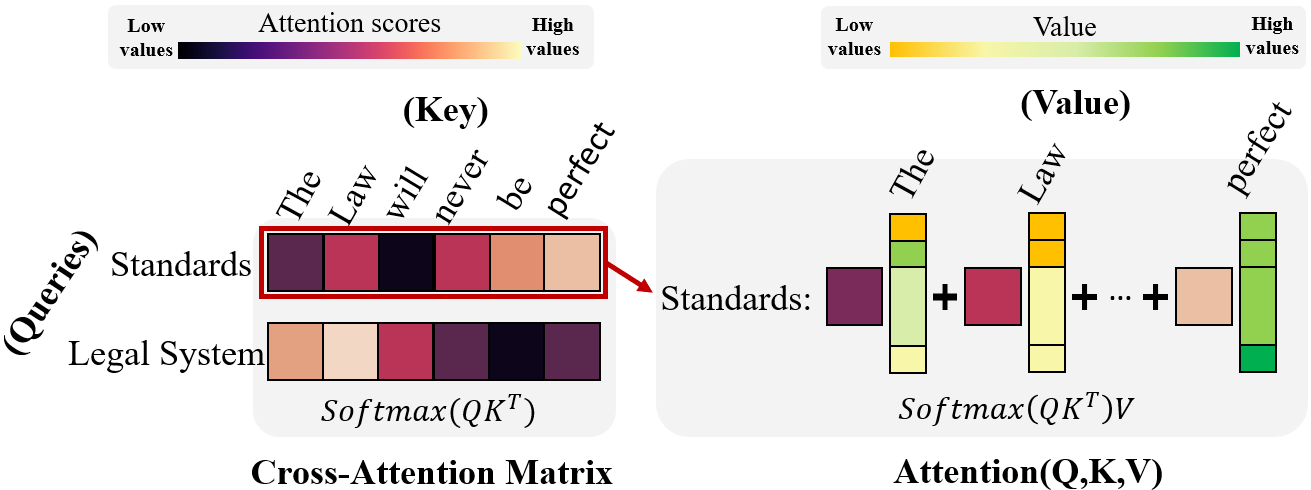}
  \caption{Example of the cross-attention mechanism: The Cross-Attention matrix quantifies the importance of a Key for a given Query, which is used to calculate a weighted average of the Value.}
  \label{fig:fig_5}
\end{figure}

Essentially, the Cross-attention mechanism works on the same principle as the Self-attention, but considering two independent sentences. These sentences are represented as the Query and the Key, respectively, while the Value is obtained from the same sentence used in the Key (Figure \ref{fig:fig_5}). The sentence used in the Query defines the criteria that the model uses to select important information, and the second sentence (Key and Value) is the information source. Accordingly, the Query is also called the context or discourse that defines the importance of the parts of a document. For example, in Figure 5, the discourses “Standards" and “Legal System” lead to a focus on different parts of the sentence.

\subsection{Proposed Technical Formulation: Discourse-based Word Probability}\label{sec3_2}
In the present section, we propose that attention scores can be formulated as a word probability in a discourse defined by the Query.

First, according to the studies \cite{mnih_2007,arora_2017}, the probability that a given word $w_{a}$ is generated under a given discourse word $w_{b}$ is proportional to the exponential of the inner product of their embedding representations (Equation \ref{eq:prob_dot_product}).
\begin{equation}
\label{eq:prob_dot_product}
    Pr\left ( w_{a} | w_{b} \right )\propto exp\left ( e_{a} \cdot e_{b}^{T} \right )
\end{equation}

From this equation, we can estimate the probability of each word $w_a$ in the sentence $A=\{w_{a}^{1},w_{a}^{2},…\}$ to be generated under the single context word $w_{b}$ as shown in Equation \ref{eq:prob_dot_product_gen}.
\begin{equation}
\label{eq:prob_dot_product_gen}
    Pr\left ( w_{a} \in A| w_{b} \right )= [ k_{1}exp{\left ( e_{a} \cdot e_{b}^{T} \right )},  k_{2}exp\left ( e_{a} \cdot e_{b}^{T} \right ),..]
\end{equation}

By assuming a common proportional constant ($k_{1}=k_{2}=...$), we can represent Equation \ref{eq:prob_dot_product_gen} as the Softmax of the Matrix product between the set of embeddings $E_a = [e_{a}^{1},e_{a}^{2},..]$ and the discourse embedding $e_{b}$, as shown in Equation \ref{eq:probability_cross} ($\varphi$ preserves the proportional constant).
\begin{equation}
\label{eq:probability_cross}
    Pr\left ( w_{a}\in A|w_{b} \right )=Softmax\left ( \frac{e_{b}\cdot {E_{a}}^{T}}{\varphi} \right )
\end{equation}

\begin{figure}[h]
  \centering
  \includegraphics[height=1.9in, width=4.8in]{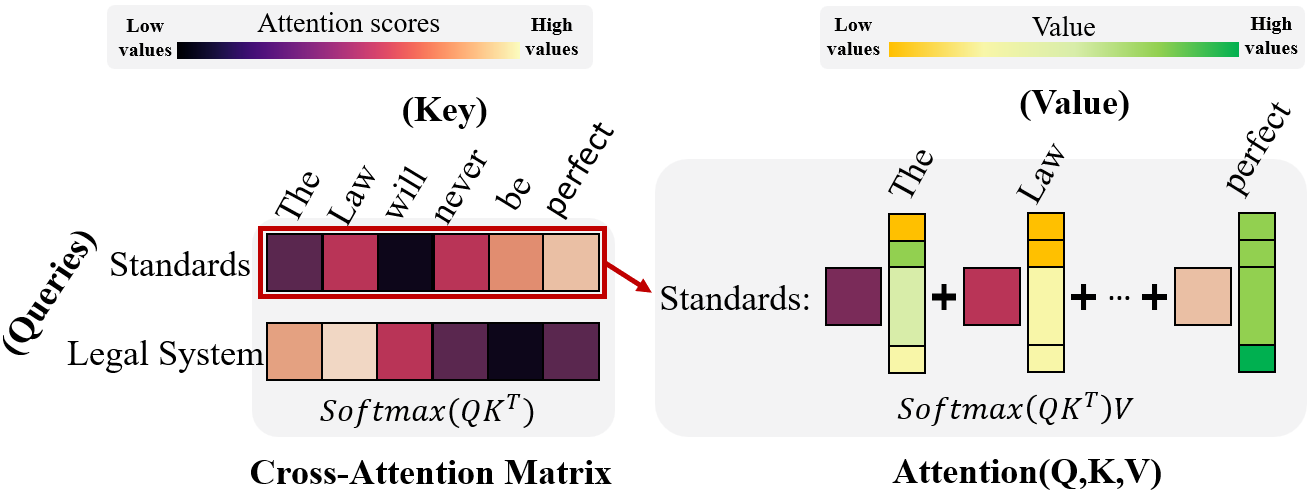}
  \caption{The Cross-Attention matrix weights can be interpreted as the probability of the Key-Value under the discourse defined by the Query.}
  \label{fig:fig_6}
\end{figure}

This equation can be interpreted as the Cross-attention Matrix between the Query $e_{b}$ and the Key $E_{a}$ (Figure \ref{fig:fig_6}). Accordingly, the values of the Cross-attention weight Matrix quantify the probability of a word from the Key sentence to be generated under a context word (Query). Similarly, the Cross-Attention mechanism generates an embedding vector that preserves the information of the words with the highest probabilities in the selected discourse.

\section{Proposed Model}\label{sec4}
\begin{figure}[h]
  \centering
  \includegraphics[height=1.8in, width=4.0in]{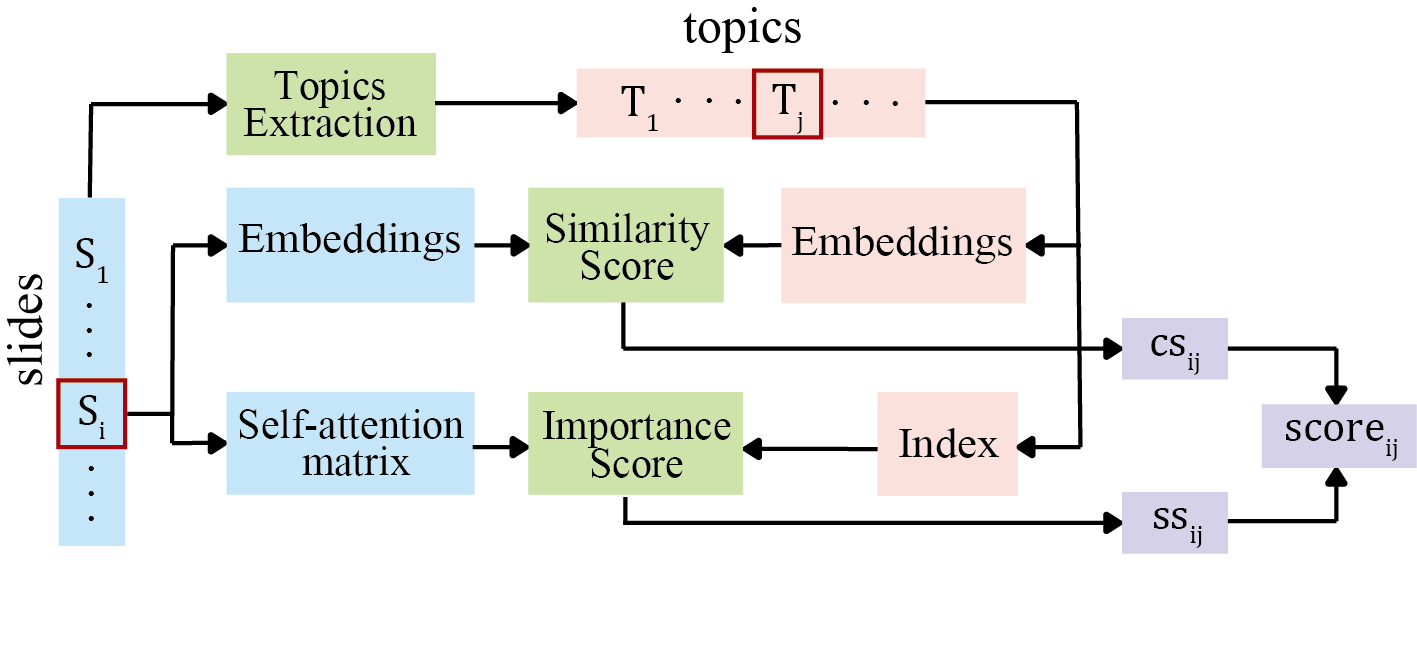}
  \caption{Overview of our proposed model: The slides are processed by the Topic Extraction module to obtain a set of topics. Then, the slide $i$ and topic $j$ are processed by the Similarity and Importance Score modules, obtaining a single score that quantitatively represents its relationship.}
  \label{fig:fig_7}
\end{figure}

LECTOR extracts a set of topic candidates from the lecture slides and assigns a unique score to each slide-topic pair (Figure \ref{fig:fig_7}). This score is defined as a linear combination of two scores, one based on the words’ importance (Importance score) and the other on the similarity between the topic and the slide embeddings (Similarity score). In this section, we will detail the different components of this system.

\subsection{Topics Extraction}\label{sec4_1}
LECTOR considers a topic candidate as an observable entity that can be represented by a keyphrase. For documents written in English, common methods use the Part-Of-Speech to extract noun phrases, which are then defined as the keyphrase candidates. In our case, the slides are written in Japanese, where the grammatical categories are more difficult to find. Consequently, we used a Bi-LSTM pretrained model to identify the nouns. Then, we defined single nouns (e.g., function, recursion) and n-gram sequences of nouns (e.g., function definition, list processing) as the topic candidates. We considered $n=2$ because longer noun phrases in Japanese tend to represent multi-topic ideas. Although the extracted set of possible topics includes all nouns and 2-gram nouns written on the slides, LECTOR can estimate the importance of these topics in the last module (Figure \ref{fig:fig_7}), and consequently, the unimportant topics are easily filtered out at the end of the model.

\subsection{Embeddings and Self-Attention Matrix Extraction}\label{sec4_2}
\begin{figure}[h]
  \centering
  \includegraphics[height=2.2in, width=4.0in]{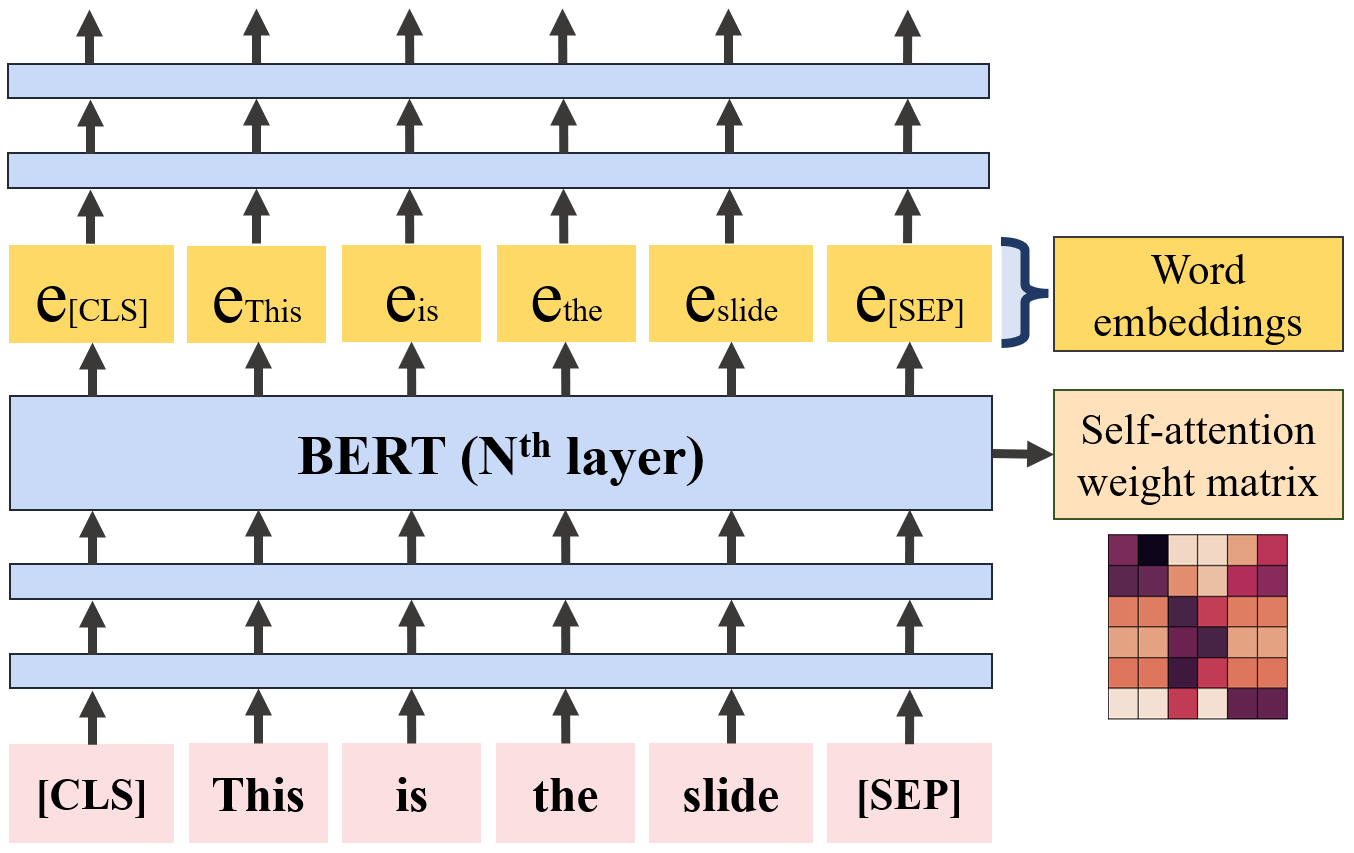}
  \caption{The word embeddings and the self-attention matrix are outputs obtained by the BERT model.}
  \label{fig:fig_8}
\end{figure}

As shown in Figure \ref{fig:fig_7}, the “Similarity score” module requires embedding vector representations of both slides and embeddings, and the “Importance score” module requires information about the Self-attention matrix of the slide. This information is extracted by a BERT model, that processes the text of lecture slides sequentially (Figure \ref{fig:fig_8}), slide by slide (word embeddings are estimated from the context of their corresponding slide). Since BERT uses subword tokenization, the extracted Self-Attention Matrix was corrected from a set of subword token-wise weights into a set of word-wise weights by using the methodology introduced by the previous work \cite{clark_2019}.

\begin{figure}[h]
  \centering
  \includegraphics[height=2.45in, width=4.8in]{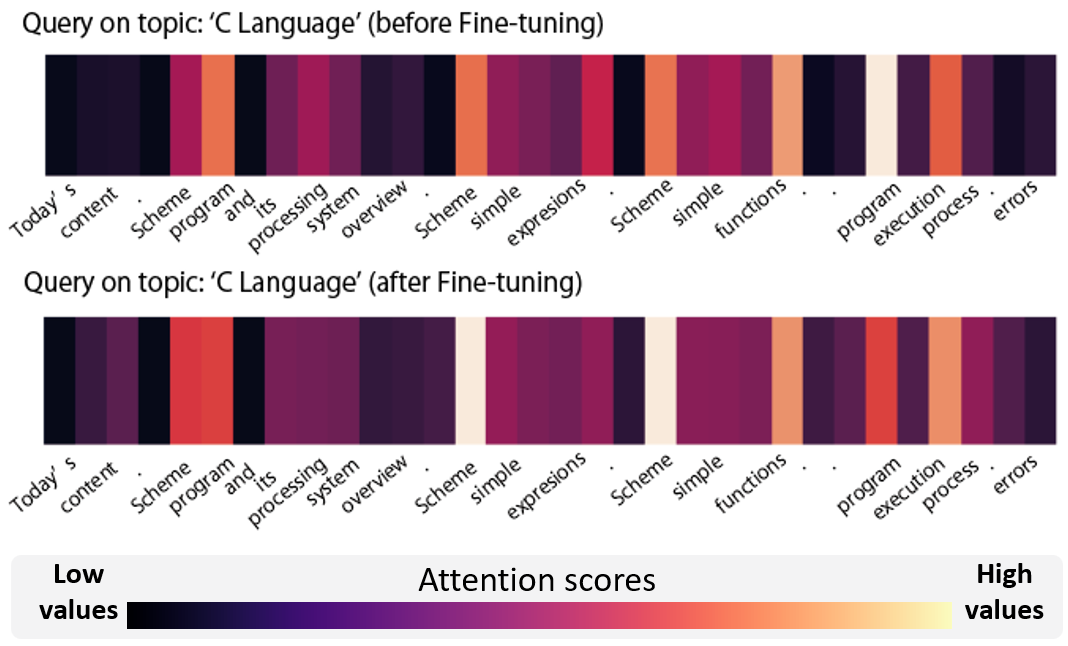}
  \caption{Differences in attention values before and after the domain-specific pre-training process: In the second case the model can better understand domain-specific relationships.}
  \label{fig:fig_9}
\end{figure}

Since BERT model is trained on a non-specific domain dataset (Wikipedia corpus), the extracted features do not encapsulate domain-specific relationships. For example, in Figure 9, the base BERT model did not clearly identify the relationship between "C Language" and "Scheme" although both are very related programming languages. For this reason, we created an additional corpus from the text of all the lecture slides of the course and used it to pre-train our BERT model on this new corpus with the Masked Language Model task \cite{devlin_2019}. As shown in Figure \ref{fig:fig_9}, after this process, the domain-specific pre-trained BERT manages to understand the relationship between "C Language" and "Scheme".

After processing all the lecture slides of a course, we collect the Self-attention Matrix $A^i$ and the matrix of word embeddings $E^i$ for each processed lecture slide $s_i$. In addition, we extract the topic embeddings $E_t^j$ from the matrix of word embeddings $E^i$ of the slides where the topic is written. Since a topic may appear on different slides, we collect the different embedding vectors and pass this set into the “Similarity score” module.

Based on previous works \cite{sun_2020, ding_luo_2021}, the second input of the “Similarity score” module, the slide vector embedding $P_s^i$, is estimated as a weighted average of its corresponding word embeddings $E^i$ extracted by our BERT model (Equation \ref{eq:sentence_embedding}).
\begin{equation}
\label{eq:sentence_embedding}
    P_{s}^{i}=\sum_{w\in s_{i}}Weight\left ( w \right )E_{w}^{i}
\end{equation}

In the literature, these weights are defined under different assumptions. The SIFRank model assumes that all words in a document are generated under a single discourse, and the AttentionRank model assumes that a document is generated in the context of a keyphrase, which in turn is generated in the context of the document.

LECTOR makes two assumptions to contextualize the information from the slide contents. Inspired by the works \cite{wang_sumiya_2010,atapattu_2017}, our assumptions leverage the lecture slides' hierarchical structures to filter out the intrinsic noise from them. These assumptions are:
\begin{itemize}
    \item The main title of a lecture material introduces the main discourse of the material. 
    \item The title of any slide introduces the specific discourse of the slide.
\end{itemize}
To formalize these assumptions as mathematical expressions, we use our proposed Discourse-based probability (Preliminaries section). Since the context presented by the titles contains different words, we generalize Equation \ref{eq:probability_cross} to a phrasal context by using the approach “Attention over attention” \cite{cui_2017}.
\begin{equation}
\label{eq:generalized_probability}
    Pr\left ( w_{a} \in A|w_{b} \in B \right )={AVE_{row}}\left ( SOF_{col}\left ( \frac{E_{b}\cdot {E_{a}}^{T}}{\varphi} \right ) \right )SOF_{row}\left ( \frac{E_{b}\cdot {E_{a}}^{T}}{\varphi} \right )
\end{equation}
where,
\begin{description}
    \item[$AVE_{row}$:] Average along the row axis.
    \item[$SOF_{col}$:] Softmax along the column axis.
    \item[$SOF_{row}$:] Softmax along the row axis.
\end{description}

\begin{figure}[h]
  \centering
  \includegraphics[height=2.4in, width=3.2in]{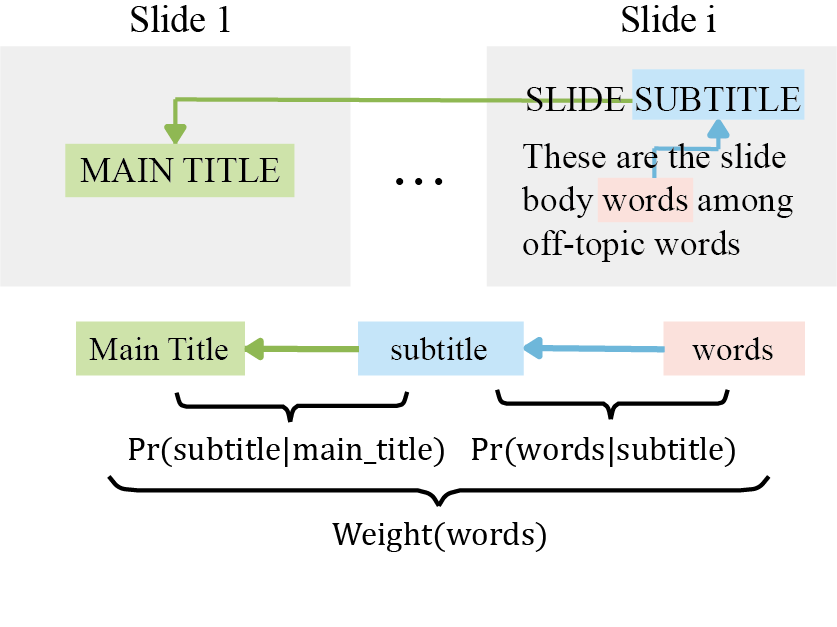}
  \caption{The process of calculating the weights: A weight quantifies the coherence of a word within the context provided by the main title and the current slide subtitle.}
  \label{fig:fig_10}
\end{figure}

Thus, given the set of words $w_t$ from the title $st_i$ and body $sb_i$, and their corresponding embeddings $E_{st}^i$ and $E_{sb}^i$, the Weights of Equation \ref{eq:sentence_embedding} are calculated as the matrix product shown in Equation \ref{eq:weight_definition}. This equation considers the probability of a word in the context introduced by the slide title $Pr\left ( w_{t}\in sb_{i}|E_{st}^{i} \right )$ and the probability of the slide title in the context introduced by the main title of the lecture $ Pr\left ( w_{t}\in st_{i}|E_{st}^{1} \right ) $. This mechanism is visualized in Figure \ref{fig:fig_10}.
\begin{equation}
\label{eq:weight_definition}
    Weight=Pr\left ( w_{t}\in st_{i}|E_{st}^{1} \right )Pr\left ( w_{t}\in sb_{i}|E_{st}^{i} \right )
\end{equation}
where,
\begin{equation}
\label{eq:aux_1}
    Pr\left ( w_{t}\in st_{i}|E_{st}^{1} \right ) = {AVE_{row}}\left ( SOF_{col}\left ( \frac{E_{st}^{1}\cdot {E_{st}^{i}}^{T}}{\varphi} \right ) \right )SOF_{row}\left ( \frac{E_{st}^{1}\cdot {E_{st}^{i}}^{T}}{\varphi} \right )
\end{equation}
\begin{equation}
\label{eq:aux_2}
    Pr\left ( w_{t}\in sb_{i}|E_{st}^{i} \right ) = Softmax\left ( \frac{E_{st}^{i}\cdot {E_{sb}^{i}}^{T}}{\varphi} \right )
\end{equation}

\subsection{LECTOR’s Importance Score}\label{sec4_3}
The score calculated by this module considers the semantic importance of the topics in the context of their corresponding slides, similar to the “Accumulated Self-Attention Calculation” module of the AttentionRank model \cite{ding_luo_2021}. As shown in Equation \ref{eq:self-attention summation}, this module quantifies the attention $a_{ij}$ that the words w belonging to a given topic $t_j$ (e.g., the words “recursion” and “function” from the topic “recursion function”) receive from all the other words $w'$ within the slide $s_i$ by summing the different weights of the Matrix $A^i$. It is important to note that this sum does not consider the attention a word receives from itself.
\begin{equation}
\label{eq:self-attention summation}
    a_{ij}=\sum_{w\in t_{j}}\sum_{w'\in s_{i}\setminus \left \{ w \right \}}A_{w'w}^{i}
\end{equation}
Since this score $a_{ij}$ is strongly influenced by the frequency of the topic's words $f_j$, the importance score ($ss_{ij}$) is calculated by correcting the attention scores with the Smooth Inverse Frequency method \cite{arora_2017}, as shown in Equation \ref{eq:sif_weights}. Here, the values of $k$ are suitable in $[10^{-4},10^{-3}]$ according to \cite{sun_2020}. We also empirically validated these values before conducting our experiments.
\begin{equation}
\label{eq:sif_weights}
    ss_{ij}=a_{ij}\left ( \frac{k}{k+f_{j}} \right )
\end{equation}

\subsection{LECTOR’s Similarity Score}\label{sec4_4}
The score calculated by this module estimates the similarity between each topic and each slide. As shown in Equation \ref{eq:cosine_similarity}, given the slide $s_i$ and the topic $t_j$, their similarity $b_{ij}$ in the embedding space is calculated by the cosine similarity between their corresponding embedding vectors ($P_s^i$ and $E_t^j$ respectively). Thanks to the contextualization of the information used to estimate the embedding vector $P_s^i$ (Equations \ref{eq:sentence_embedding} and \ref{eq:weight_definition}), this representation preserves the most important information of each slide.
\begin{equation}
\label{eq:cosine_similarity}
    b_{ij}=\frac{P_{s}^{i} \cdot E_{t}^{j}}{||P_{s}^{i}||\: ||E_{t}^{j}||}
\end{equation}
Since a topic is sometimes found in different contexts (different slides), we calculate a set of similarities $b_ij$ for the set of $E_t^j$ embedding vectors collected from the different slides. Finally, as shown in Equation \ref{eq:similarity_score}, the Similarity score $cs_{ij}$ is calculated as the mean of this set of values $b_{ij}$ between the slide $s_i$ embedding and all the instances of the topic $t_j$ across all different contexts of the same word. Since this average score sometimes gives relatively high importance to infrequent topics that are irrelevant, we also considered a softened value of frequency.
\begin{equation}
\label{eq:similarity_score}
cs_{ij}=\left ( \frac{1}{f_{j}}\sum_{topic j} b_{ij}\right )f_{j}^{\alpha },\alpha\in \left [0,0.25 \right ]
\end{equation}

\subsection{LECTOR’s Final Score}\label{sec4_5}
The final score for the pair of topic $t_j$ and slide $s_i$ is a linear combination of the importance and the similarity scores (previously normalized), as shown in Equation 14. Here, the parameter d defines the importance the model gives to each score value. Finally, these scores are outputted as the matrix M, where the element  $M_{ij}=score_{ij}$.
\begin{equation}
\label{eq:final_score}
    score_{ij}=d*ss_{ij}+\left ( 1-d \right )*cs_{ij}
\end{equation}

\subsection{LECTOR and Reading Logs Integration}\label{sec4_6}
LECTOR can be integrated into traditional feature extraction methods by multiplying its output M with the slide-wise reading logs. From this integration, we can design new features that consider slide content to provide new insights into students' reading behavior. In the current work, we estimated students' topic preferences and evaluated their benefits in extending at-risk students’ predictive models.

To derive the vector of topic preferences, we used the Reading Time students spent on different slides. For a given student, we created an n dimensional vector containing their reading time on each slide (n is the number of slides in the material). We then applied the $l_1$-normalization to this vector. This process ensures that the elements of the vector quantify the percentage of engagement with the different slides, regardless of the total time the student dedicates to the material.

\begin{figure}[h]
  \centering
  \includegraphics[height=1.4in, width=4.8in]{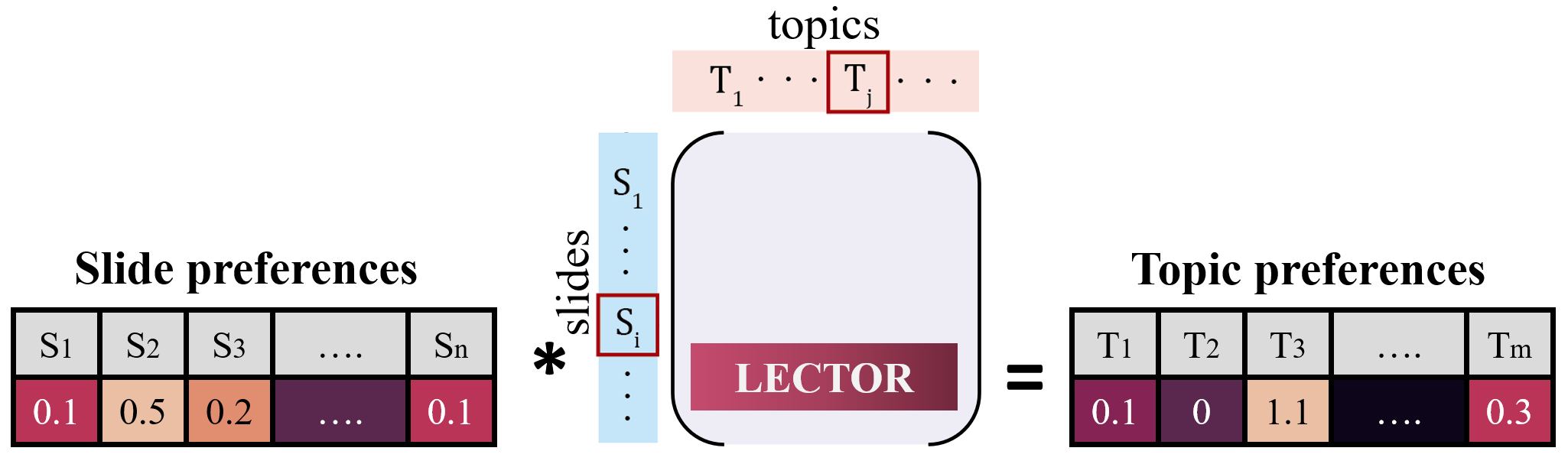}
  \caption{The vector of topic preferences is estimated by multiplying the vector of Slide preferences by the matrix of Slide-Topic relationships estimated by LECTOR.}
  \label{fig:fig_11}
\end{figure}

Finally, the vector of topic preferences is given by the matrix product between the vector of slide preferences and the matrix M estimated by LECTOR (Figure \ref{fig:fig_11}). The resulting vector has m dimensions (m is the number of topics considered), and its elements take high values when the student prefers slides that are highly related to the corresponding topic. 

\section{Experimental Setting}\label{sec5}

\subsection{Dataset}\label{sec5_1}
\begin{figure}[h]
  \centering
  \includegraphics[height=2.4in, width=4.2in]{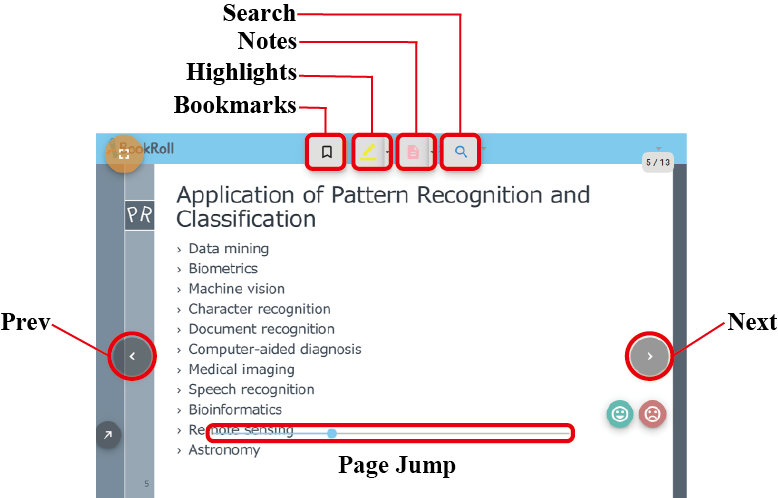}
  \caption{User Interface of the BookRoll system used in our experiments.}
  \label{fig:fig_bookroll}
\end{figure}

The data was collected from the BookRoll system, which tracks student interactions with digital learning materials. As shown in Figure \ref{fig:fig_bookroll}, BookRoll provides a simple interface that allows students to navigate through lecture slides using \textit{Next}, \textit{Previous}, \textit{Jump}, and \textit{Search} functions. It also allows students to highlight content, add notes, and bookmark pages. These interactions are logged as reading event streams (reading logs), which include student IDs, action types, and timestamps, as summarized in Table \ref{table_tadd1}. Traditionally, these logs are processed into a feature vector that summarizes the frequency of students' reading activities along with their reading time (Table \ref{table_tadd2}).

\begin{table}[h]
\centering
\caption{Example of reading logs collected by BookRoll.}
\label{table_tadd1}
\begin{tabular}{ccccc}
\hline
\textbf{User\_ID} & \textbf{Material\_ID} & \textbf{Operation} & \textbf{Page N.} & \textbf{Event\_Time} \\ \hline
A20XX\_U1 & A20XX\_C1 & PREV   & 5 & A/B/20XX 15:12:09 \\
A20XX\_U1 & A20XX\_C1 & NEXT   & 4 & A/B/20XX 15:12:52 \\
A20XX\_U2 & A20XX\_C1 & JUMP   & 2 & A/B/20XX 15:13:06 \\
A20XX\_U1 & A20XX\_C1 & NEXT   & 5 & A/B/20XX 15:13:25 \\
A20XX\_U1 & A20XX\_C1 & MARKER & 6 & A/B/20XX 15:13:37 \\
A20XX\_U1 & A20XX\_C1 & NEXT   & 6 & A/B/20XX 15:13:55 \\
A20XX\_U2 & A20XX\_C1 & PREV   & 9 & A/B/20XX 15:14:11 \\
A20XX\_U2 & A20XX\_C1 & PREV   & 8 & A/B/20XX 15:14:41 \\
A20XX\_U2 & A20XX\_C1 & PREV   & 7 & A/B/20XX 15:15:13 \\ \hline
\end{tabular}
\end{table}

\begin{table}[h]
\centering
\caption{Set of features traditionally used to represent students' reading activities \cite{chen_2021}.}
\label{table_tadd2}
\begin{tabular}{ll}
\hline
\multicolumn{1}{c}{\textbf{Feature}} & \multicolumn{1}{c}{\textbf{Description}} \\ \hline
OPEN                                 & Opening a material                       \\
CLOSE                                & Closing a material                       \\
SEARCH                               & Searching a keyword                      \\
NEXT                                 & Moving to a next page                    \\
PREV                                 & Moving to a previous page                \\
PAGE\_JUMP                           & Jumping to another page                  \\
ADD\_BOOKMARK                        & Attaching a bookmark                     \\
BOOKMARK\_JUMP                       & Jumping to a bookmark page               \\
DEL\_BOOKMARK                        & Removing a bookmark                      \\
ADD\_MARKER                          & Highlighting text                        \\
DEL\_MARKER                          & Deleting a highlighted area              \\
ADD\_MEMO                            & Writing a note                           \\
DEL\_MEMO                            & Removing a note                          \\
CHANGE\_MEMO                         & Modifying a note                         \\
READ\_TIME                           & Total time spent on the system           \\ \hline
\end{tabular}
\end{table}

\begin{figure}[h]
  \centering
  \includegraphics[height=2.8in, width=5.0in]{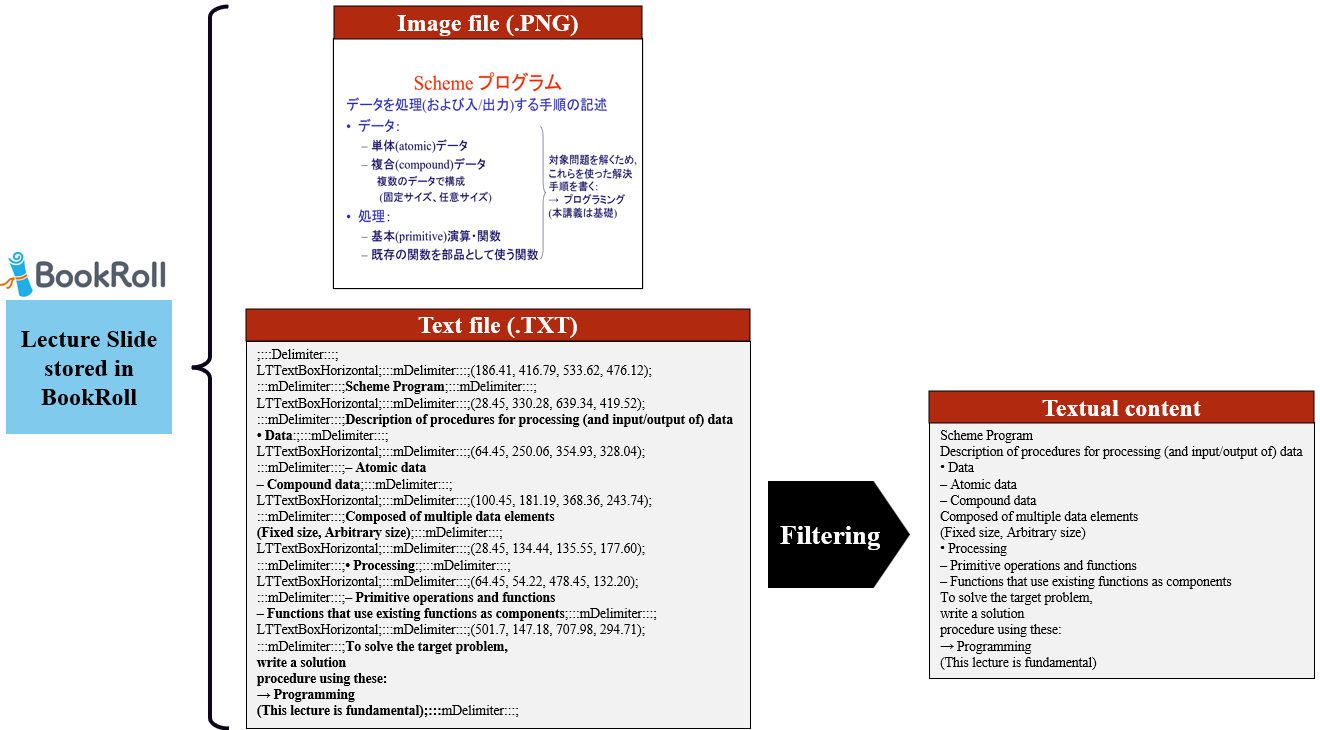}
  \caption{A reading material slide is stored in BookRoll as a PNG file and TXT file.}
  \label{fig:fig_bookroll_in}
\end{figure}

In addition to reading logs, BookRoll stores the content of uploaded materials. As shown in Figure \ref{fig:fig_bookroll_in}, each lecture slide is stored as a PNG image and as structured textual data in a corresponding TXT file (extracted by an OCR system built into BookRoll). The textual data used in this study was retrieved from these TXT files after a filtering process that removes box spatial information, noise characters, and unnecessary spaces (Figure \ref{fig:fig_bookroll_in}).

Based on this system, we constructed two datasets corresponding to two different modalities from the same learning environment. The first dataset consists of the textual content of 2255 slides from 70 e-book materials delivered in a “Programming Theory” course from 2019 to 2023. The second dataset consists of anonymized reading logs of 218 students who used BookRoll in the mentioned courses, excluding the 2023 cohort due to ongoing anonymization procedures at the time of data collection (Table \ref{table_dataset}).
\begin{table}[h]
\centering
\caption{Description of the Dataset courses.}
\label{table_dataset}
\begin{tabular}{lccccc}
\hline
\multicolumn{1}{c}{\multirow{2}{*}{\textbf{Course}}} &
  \textbf{Number} &
  \textbf{Number of} &
  \textbf{Number of} &
  \multicolumn{1}{c}{\textbf{Number of}} &
  \textbf{Number of} \\
\multicolumn{1}{c}{} &
  \textbf{of slides} &
  \textbf{materials} &
  \textbf{students} &
  \multicolumn{1}{c}{\textbf{logs}} &
  \textbf{weeks} \\ \hline
A-2019 & 620 & 22 & 52 & 130,239 & 8 \\
A-2020 & 422 & 12 & 60 & 142,754 & 7 \\
A-2021 & 423 & 12 & 54 & 130,330 & 8 \\
A-2022 & 400 & 12 & 52 & 197,389 & 7 \\
A-2023 & 390 & 12 & -  & -       & - \\ \hline
\end{tabular}
\end{table}

The courses were offered by the School of Engineering of Kyushu University for 7 to 8 weeks, following the same curricula. In these courses, students accessed lecture material slides at their preferred time using the e-book system Bookroll \cite{ogata_2017,flanagan_2018}. The evaluation system consisted of weekly reports on programming exercises and a final test. At the end of the course, students were assigned letter grades (A, B, C, D, and F). At-risk students were defined as those who received a grade of D or F, consistent with previous studies. Due to university privacy policies, no demographic data, such as gender, race, or socioeconomic background, were collected. However, available records indicate that students were between 19 and 22 years old, and they were fluent in Japanese, the language of instruction.

\subsection{Experiments}\label{sec5_15}
We conducted two main experiments, as listed below:
\begin{description}
    \item[1. Experiment 1:] Keyphrases extraction from e-book lecture slides
    \item[2. Experiment 2:] Student reading preferences for detecting at-risk students
\end{description}

The first experiment was designed to evaluate LECTOR's ability to extract the most important information from lecture slides (RQ1). We processed textual content from slides using representative NLP models, including LECTOR (independent variable), and evaluated their performance in the keyphrase extraction task (dependent variable). Through comparative analysis, we determined whether LECTOR outperformed the baseline models in this task.

The second experiment was designed to assess the benefits of integrating LECTOR with traditional reading activity data for predicting at-risk students (RQ2). Here, we processed reading logs using both traditional features and features derived from the LECTOR integration (independent variable) and evaluated their performance in predicting at-risk students (dependent variable). Through qualitative and quantitative comparisons, we determined whether the integration of LECTOR with traditional data offers advantages for predicting at-risk students.

\subsection{Evaluation and Analysis Methods}\label{sec5_2}
Despite the significantly different approaches of Experiments 1 and 2, both used Recall, Precision, and F1-scores to measure performance in prediction tasks. These values are calculated as shown in Equations \ref{eq:recall}, \ref{eq:precision}, and \ref{eq:fscore}.
\begin{equation}
\label{eq:recall}
    Recall=\frac{TP}{TP+FN}
\end{equation}

\begin{equation}
\label{eq:precision}
    Precision=\frac{TP}{TP+FP}
\end{equation}

\begin{equation}
\label{eq:fscore}
    F_{1}\cdot score=\frac{2*Recall*Precision}{Recall+Precision}
\end{equation}
where, TP: True Positives, FP: False Positives, FN: False Negatives.

These values are within the range of [0,1]. However, in unsupervised keyphrase extraction, they are usually represented as percentages (a range from 0 to 100) to highlight performance differences. Therefore, this paper uses the percentage notation.

In addition, Experiment 2 used Fisher's Discriminant Ratio (FDR) \cite{dougherty} to evaluate the discriminative power of a feature (a reading characteristic) in distinguishing two groups of students with different grades. Given two populations and a specific feature measuring one of their reading characteristics, the FDR is calculated as shown in Equation \ref{eq:fdr}.
\begin{equation}
\label{eq:fdr}
    FDR=\frac{\mu_{1}^{2}-\mu_{2}^{2}}{\sigma _{1}^{2}-\sigma _{2}^{2} }
\end{equation}
Where:
\begin{description}
    \item[$\mu_i$:] Mean of the feature measured from group $i$
    \item[$\sigma_i$:] Standard deviation of the feature measured from group $i$
\end{description}
The FDR can take any positive real number, with larger values indicating greater discriminative power of the feature to distinguish between the two populations. In simple Machine Learning models, a high FDR often translates to a better performance in predicting a new sample's group.

\section{Experiment 1: Keyphrases Extraction from E-book Lecture Slides}\label{sec6}
Our first experiment assessed the ability of LECTOR to quantitatively extract Slide-Topic relationships. Given the large number of Slide-Topic combinations and the inherent subjectivity in their relationships, a direct evaluation was impractical. Accordingly, we indirectly measured the quality of a matrix of Slide-Topic relationships by evaluating its ability to extract keyphrases. Specifically, we conducted a comparative analysis between LECTOR and previous models using two methods: the first is a technical evaluation tailored to keyphrase extraction models, and the second is a human evaluation of the extracted keyphrases.

\subsection{Technical Evaluation Formulation}\label{sec6_1}
To extract keyphrases from our models, we calculated the keyphrase score $mt_j$ for all topics $t_j$ as the sum of the scores $M_ij$ (Slide-Topic relationships) across all slides (Equation \ref{eq:exp1_score}). Accordingly, the topics with the highest scores are those strongly related to the majority of slides.
\begin{equation}
\label{eq:exp1_score}
mt_{j}=\sum_{i=1}^{\#slides}M_{ij}
\end{equation}

To evaluate a model's predictive performance, we defined $@n$ as the set of $n$ topics with the highest $mt_j$ score. Here, $@n$ represents a set of $n$ keyphrases estimated by the model. We then compared the $@n$ set with the Ground Truth (set of keyphrases from the course syllabus), measuring the Precision (P), Recall (R), and F1-score (F1) performance.

We included four baseline models. The first is the TF-IDF model \cite{salton_1988}, the predominant approach in the slide text processing literature. The second is the binary method for reader modeling proposed by \cite{wang_2022} (Figure \ref{fig:fig_2}). The third is the graph-based unsupervised keyphrase extraction model TextRank \cite{mihalcea_2004,takii_2024}. The fourth is AttentionRank \cite{ding_luo_2021}, which represents the state-of-the-art in unsupervised keyphrase extraction.

The Ground-Truth keyphrases were obtained from the syllabus of the course. This set consists of 9 keyphrases: \textit{Scheme}, \textit{Data Structure}, \textit{List Processing}, \textit{Recursion}, \textit{Expression}, \textit{Condition}, \textit{Design Recipe}, \textit{Function}, \textit{High-level function}.

\subsection{Technical Evaluation Results}\label{sec6_2}
 
\begin{table}[h]
\centering
\caption{Summary of the models' performances in the Technical evaluation. Evaluations are conducted at the top “n” keyphrases predicted by the method.}
\label{table_1}
\begin{tabular}{cllll}
\hline
\textbf{ n} & \textbf{Model} & \multicolumn{1}{c}{\textbf{P}} & \multicolumn{1}{c}{\textbf{R}} & \multicolumn{1}{c}{\textbf{F1}} \\ \hline
\multirow{5}{*}{5}    & TF-IDF              & 20.00          & 11.11          & 14.39          \\
                      & Binary score        & 20.00          & 11.11          & 14.39          \\
                      & TextRank            & 20.00          & 13.33          & 17.25          \\
                      & AttentionRank       & \textbf{24.00} & \textbf{13.33} & \textbf{17.25} \\
                      & LECTOR              & 20.00          & 11.11          & 14.39          \\ \hline
\multirow{5}{*}{10}   & TF-IDF              & 16.00          & 17.78          & 16.94          \\
                      & Binary score        & 12.00          & 13.33          & 12.73          \\
                      & TextRank            & 10.00          & 11.11          & 10.62          \\
                      & AttentionRank       & \textbf{30.00} & \textbf{33.33} & \textbf{31.68} \\
                      & LECTOR              & \textbf{30.00} & \textbf{33.33} & \textbf{31.68} \\ \hline
\multirow{5}{*}{15}   & TF-IDF              & 20.00          & 33.33          & 25.11          \\
                      & Binary score        & 20.00          & 33.33          & 25.11          \\
                      & TextRank            & 6.67           & 11.11          & 8.44           \\
                      & AttentionRank       & 20.00          & 33.33          & 25.11          \\
                      & LECTOR              & \textbf{26.67} & \textbf{44.44} & \textbf{33.44} \\ \hline
\multirow{5}{*}{Best} & TF-IDF (n=14)       & 24.00          & 35.55          & 28.48          \\
                      & Binary score (n=13) & 23.85          & 33.33          & 27.90          \\
                      & TextRank (n=18)     & 10.78          & 31.11          & 15.96          \\
                      & AttentionRank (n=8) & \textbf{35.83} & 33.33          & 34.61          \\
                      & LECTOR (n=12)       & 33.01          & \textbf{44.44} & \textbf{37.92} \\ \hline
\multirow{5}{*}{Mean} & TF-IDF              & 12.03          & 45.85          & 15.86          \\
                      & Binary score        & 11.53          & 44.40          & 15.16          \\
                      & TextRank            & 8.57           & 37.91          & 11.93          \\
                      & AttentionRank       & 11.96          & 40.20          & 15.06          \\
                      & LECTOR              & \textbf{14.91} & \textbf{57.73} & \textbf{20.14} \\ \hline
\end{tabular}
\end{table}

\begin{table}[h]
\centering
\caption{Most important topics of each model. ENG: originally written in English.}
\label{table_2}
\begin{tabular}{clllll}
\hline
\textbf{n} &
  \multicolumn{1}{c}{\textbf{Tf-idf}} &
  \multicolumn{1}{c}{\textbf{Binary score}} &
  \multicolumn{1}{c}{\textbf{TextRank}} &
  \multicolumn{1}{c}{\textbf{AttentionRank}} &
  \multicolumn{1}{c}{\textbf{LECTOR}} \\ \hline
1  & function     & function     & function     & function        & function            \\
2  & list         & list         & thing        & example problem & data                \\
3  & list (ENG)   & define (ENG) & example      & definition      & list                \\
4  & i (ENG)      & definition   & definition   & recursion       & definition          \\
5  & define (ENG) & cond (ENG)   & explanation  & example         & program             \\
6  & definition   & data         & for          & value           & computation         \\
7  & page         & list (ENG)   & define (ENG) & define (ENG)    & function definition \\
8  & data         & empty        & computation  & expression      & expression          \\
9  & count        & count        & value        & argument        & example problem     \\
10 & program      & i (ENG)      & case         & computation     & recursion           \\
11 & value        & value        & possible     & list            & data definition     \\
12 & expression   & expression   & method       & else (ENG)      & list processing     \\
13 & cond (ENG)   & recursion    & number       & empty (ENG)     & program design      \\
14 & example      & else (ENG)   & can be       & element         & recursion function  \\
15 & recursion    & element      & result       & count           & exercises           \\ \hline
\end{tabular}
\end{table}

\begin{table}[h]
\centering
\caption{Comparison of F1-scores for keyphrase extraction on benchmark datasets and a lecture slides dataset.}
\label{table_tadd3}
\begin{tabular}{l|ccc|ccc}
\hline
\multicolumn{1}{c|}{\multirow{2}{*}{}} & \multicolumn{3}{c|}{\textbf{General Documents}} & \multicolumn{3}{c}{\textbf{Lecture Slides}} \\
\multicolumn{1}{c|}{}                  & \textbf{@5}   & \textbf{@10}   & \textbf{@15}   & \textbf{@5}  & \textbf{@10}  & \textbf{@15} \\ \hline
TF-IDF                                 & 11.91         & 13.59          & 13.87          & 14.39        & 16.94         & 25.11        \\
AttentionRank                          & 24.02         & 33.26          & 36.35          & 17.25        & 31.68         & 25.11        \\ \hline
\end{tabular}
\end{table}

The results are summarized in Table \ref{table_1}. Here, "n" represents the number of keyphrases considered in the evaluation, "Best" is the best performance of the model among all "n" evaluations, and "Mean" is the average of the performances within the first @100 evaluations. For further reference, Table \ref{table_2} details the keyphrases extracted by the different models at $n = 15$.

At $n=5$, all models correctly identified the keyphrase \textit{function}, but AttentionRank also identified the keyphrase \textit{recursion}. Despite its lower performance, LECTOR identifies the topics \textit{“data”} and \textit{“list”}, which are also important (related to the syllabus keyphrases \textit{“data structure”} and \textit{“list processing”}). In contrast, statistical models' sets include frequent terms such as \textit{“i"}, \textit{“cond”} and \textit{“define”} (from code examples) and TextRank includes very general topics such as \textit{“thing”} and \textit{“explanation”}. These last results suggest that limiting the information to the frequency and co-occurrence of topics (methods predominant in current educational developments) may give high importance to the wrong parts of the reading content

For values of $n$ higher than $10$, attention-based models outperform other approaches. Specifically, LECTOR achieves the highest performance at $@15$, “Best” and “Mean” with F1-scores of 33.44\%, 37.92\%, and 20.14\%, respectively. These results show that AttentionRank struggles to identify new keyphrases, whereas LECTOR does not. As shown in Table \ref{table_2}, this difficulty arises because AttentionRank tends to include common programming terms such as \textit{“define”}, \textit{“else”}, and \textit{“empty”} in its $@15$ set, which degrades its performance.

Interestingly, when comparing these results to those reported on general narrative-style datasets, we observe that the difference in performance varies significantly across models rather than showing a uniform decrease.

\begin{itemize}
\item TF-IDF, despite being a simple statistical model, achieves higher performance on lecture slides than on general documents (Table \ref{table_tadd3}). These results depict TF-IDF's robustness and generalizability across domains, potentially explaining its widespread adoption in educational tools.
\item In contrast, AttentionRank shows a more pronounced drop in performance compared to its results on general NLP datasets (Table \ref{table_tadd3}). This suggests that recent advances in unsupervised keyphrase extraction may not adapt well to educational contexts, especially for extracting key concepts from lecture slides.
\end{itemize}

\subsection{Human Evaluation Formulation}\label{sec6_3}
Our previous results show the potential of LECTOR to improve existing educational tools, which are primarily based on the TF-IDF model. The Human Evaluation extends these findings by collecting feedback from students on keyphrases extracted by LECTOR and the TF-IDF model (baseline). In this evaluation, we extracted the $@5$ top keyphrases from each slide for both models and combined them into a list of keyphrase candidates. Students were then presented with the slide and its corresponding list of candidates, from which they could select up to 5 keyphrases.

The students' selections were used to empirically estimate the Ground Truth values and measure the predictive performance of both models. We estimated the Ground Truth in two ways. In the first, the Ground Truth was determined by the 5 keyphrases with the highest number of student selections, and the performances were calculated at a Slide-Level. In the second case, the Ground Truth was defined directly by each student's selection, and the performances were calculated at a Student-Level.

This evaluation was conducted on the textual content of course A-2023 (Table 1). The evaluators consisted of a group of 28 students who had completed this course. Since the evaluation of all 390 slides in the dataset required approximately 8 hours of work, the students were divided into two groups (15 and 13 students), with each group evaluating half of the total number of slides.

\subsection{Human Evaluation Results}\label{sec6_4}
\begin{figure}[h]
  \centering
  \includegraphics[height=1.2in, width=5.0in]{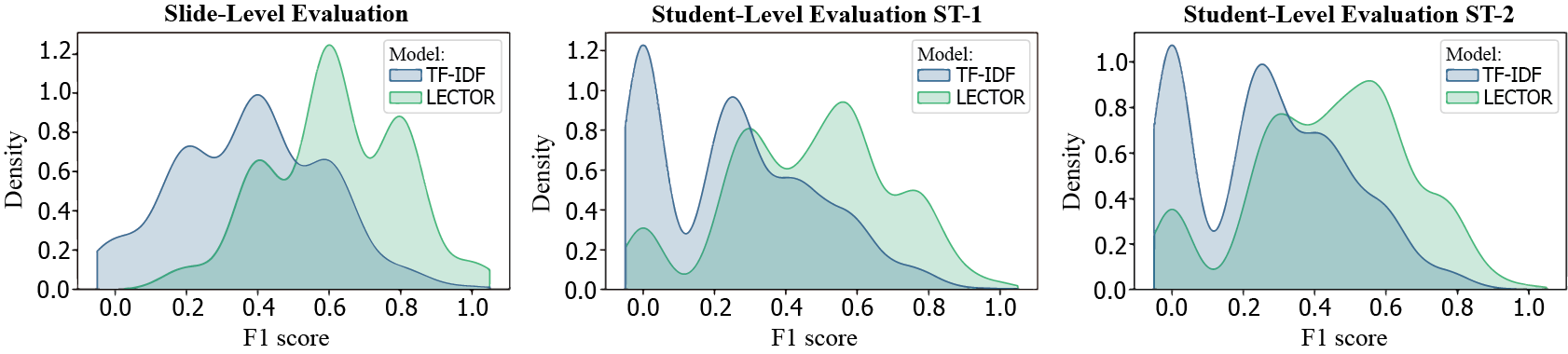}
  \caption{LECTOR and TF-IDF F1-score distributions in the three considered evaluations.}
  \label{fig:fig_12}
\end{figure}

The results are summarized in Table \ref{table_1_2_1}. The "GT" column indicates the method used to estimate the Ground Truth values. "SL" represents the Slide-Level approach, while "ST" represents the Student-Level approach. The Student-Level estimation has two cases, ST1 and ST2, which correspond to the two groups of students who each evaluated half of the total slides. Figure \ref{fig:fig_12} complements the tabular data by visualizing the distributions of the F1-scores of the models in the three evaluations.

\begin{table}[h]
\centering
\caption{Model performance from empirically estimated Ground Truth.}
\label{table_1_2_1}
\begin{tabular}{clllllll}
\hline
\textbf{GT} & \textbf{Model} & \textbf{Precision} & \textbf{p-value} & \textbf{Recall} & \textbf{p-value} & \textbf{F-score} & \textbf{p-value} \\ \hline
\multirow{2}{*}{SL}   & TF-IDF & 37.7±20.6 & 4.7E-53  & 37.7±20.6 & 4.7E-53  & 37.7±20.6 & 4.7E-53  \\
                      & LECTOR & 61.8±18.1 &          & 61.8±18.1 &          & 61.8±18.1 &          \\ \hline
\multirow{2}{*}{ST-1} & TF-IDF & 21.9±20.4 & 1.0E-186 & 34.3±31.5 & 2.1E-269 & 25.3±22.6 & 1.8E-236 \\
                      & LECTOR & 39.5±22.6 &          & 65.3±30.5 &          & 46.5±22.9 &          \\ \hline
\multirow{2}{*}{ST-2} & TF-IDF & 23.5±20.2 & 1.9E-116 & 36.2±30.5 & 1.4E-160 & 26.9±21.7 & 7.0E-150 \\
                      & LECTOR & 38.1±21.5 &          & 61.4±30.4 &          & 44.4±22.2 &          \\ \hline
\end{tabular}
\end{table}

LECTOR consistently outperformed the TF-IDF model, with a difference validated by a T-test (390 samples in SL, 2760 samples in ST-1, and 2301 samples in ST-2). The significant difference indicates that LECTOR selections are more closely aligned with students' perceptions of relevant information on each lecture slide. This improvement in information retrieval is crucial for improving existing educational tools, such as slide summarization systems \cite{shimada_2018}, recommendation systems \cite{nakayama_2019,okubo_2020}, and students' reading behavior modeling \cite{wang_2022}. 
In the Slide-Level evaluation, LECTOR achieved an average F-score of 61.8\% across 390 slides (on average 3 of the 5 empirical keyphrases were correctly extracted). In the Student-Level evaluation, LECTOR achieved an average F-score of 45.5\%. Although this value is lower than the Slide-Level evaluation, the average Recall of 63.4\% remains comparable, indicating that students sometimes selected fewer than 5 keyphrases for each slide. This behavior is caused by two main factors: some slides contain few keyphrases, and LECTOR tends to be less effective on certain slides, especially those that are highly visual or lack clear titles.

\section{Experiment 2: Student Reading Preferences for Detecting At-risk Students}\label{sec7}
Our second experiment assessed the benefits of integrating LECTOR into traditional reading activity data. Specifically, we used students' topic preferences (Figure \ref{fig:fig_11}) to predict at-risk students, contrasting their results with traditional representations and analyzing potential contributions.

This experiment is divided into two evaluations. In the first, students' topic preferences were analyzed at a feature level to identify differences between students with different grades (Basic Separability Analysis). In the second, the features were automatically processed by a Machine Learning model to predict at-risk students (At-risk students' prediction).

\subsection{Basic Separability Analysis Formulation}\label{sec7_1}
In this evaluation, we selected different features (reading characteristics) to find those that could most effectively separate two groups of students with different grades. By evaluating the separability of the distributions in the different cases, we compared topic-based and traditional representations. The topic-based representations were given by LECTOR's vector of topic preferences (Figure \ref{fig:fig_11}), where each feature represents the relative engagement with a particular topic. On the other hand, the traditional feature (baseline) was derived from the original data source of this vector, the “Reading Time”.

As a first step, we grouped the students according to their grades (A, B, C, D, and F) and selected two groups. We then used Fisher's Discriminant Ratio (FDR) to measure the separability of their distributions for each feature. The decision to use the FDR is based on its quality to represent the discriminative power of a feature and its potential to improve predictive performance in ML models \cite{dougherty}.

From the set of topics, we selected the one with the highest FDR (Best Topic: B\_Topic). We then computed the FDR of the distributions of READING TIME and B\_Topic of the two groups of students. The obtained values showed the degree of separability in both cases.

\begin{figure}[h]
  \centering
  \includegraphics[height=1.35in, width=5.0in]{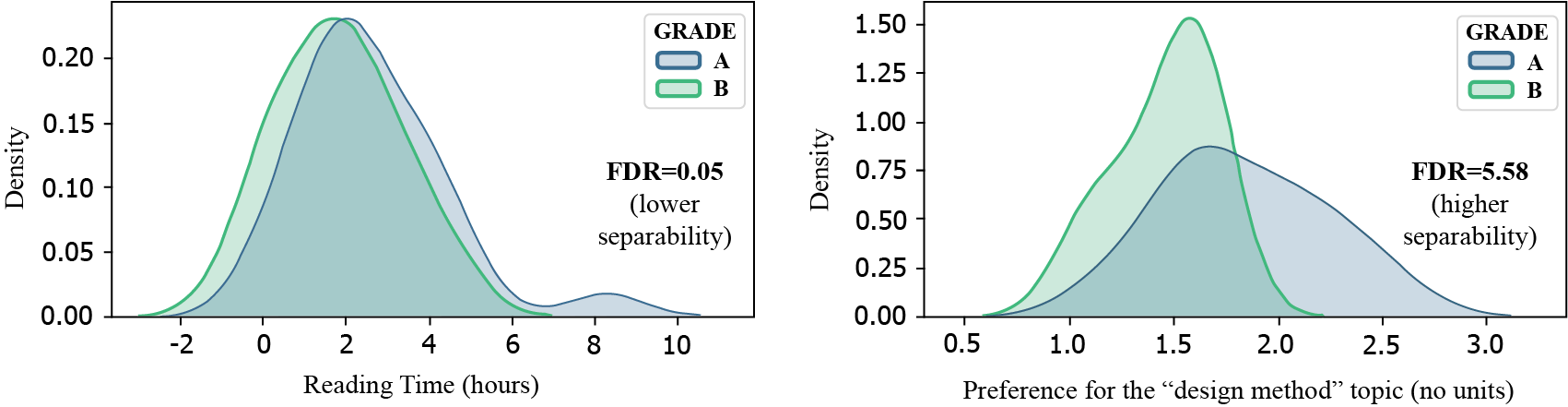}
  \caption{Populations of students with grades A and B: The distributions of their Reading time overlap, but the distributions of their preference for the topic design method are more separated.}
  \label{fig:fig_13}
\end{figure}

For example, Figure \ref{fig:fig_13} shows the distributions of READING TIME and B\_Topic (Design method) for two groups of students with grades A and B. Although both groups spent almost the same amount of time reading, the students with grade A showed a preference for the “design method” topic. In this case, our FDR analysis indicated that, for a ML model, the relative preference for the “design method” topic is a better feature than the READING TIME (FDR values of 5.58 versus 0.05).

\subsection{Basic Separability Analysis Results}\label{sec7_2}
The results are summarized in Table \ref{table_4}. This table includes five comparison cases: groups of students with consecutive grades (A-B, B-C, C-D, D-F) and at-risk/non-risk students. For reference, the example from Figure \ref{fig:fig_13} is included in the “A-B” column, “Week 2 Out-class” row. We limited our analysis to course A-2019 to avoid an overwhelming number of results. Additionally, the fifth and subsequent weeks were not considered since the data was insufficient to perform the analysis (due to students' disengagement). 

\begin{table}[h]
\centering
\caption{Separability (FDR) values obtained by a reading characteristic in distinguishing students with different grades (R.Time: Reading Time, Topic P.:Topic Preference).}
\label{table_4}
\begin{tabular}{clccccc}
\hline
\multicolumn{1}{l}{} &
   &
  \textbf{A-B} &
  \textbf{B-C} &
  \textbf{C-D} &
  \textbf{D-F} &
  \textbf{At-risk} \\ \hline
\multirow{4}{*}{\begin{tabular}[c]{@{}c@{}}WEEK 1\\ (IN-\\ CLASS)\end{tabular}} &
  R. Time &
  0.034 &
  0.262 &
  2.782 &
  0.023 &
  1.111 \\
 &
  Topic P &
  \textbf{1.452} &
  \textbf{612.4} &
  \textbf{46.86} &
  \textbf{3.233} &
  \textbf{4.325} \\
 &
  (B\_Topic) &
  (expressions) &
  (data) &
  (exercises) &
  (design &
  (execution) \\
 &
   &
  \multicolumn{1}{l}{} &
  \multicolumn{1}{l}{} &
  \multicolumn{1}{l}{} &
  method) &
   \\ \hline
\multirow{4}{*}{\begin{tabular}[c]{@{}c@{}}WEEK 1\\ (OUT-\\ CLASS)\end{tabular}} &
  R. Time &
  0.041 &
  0.002 &
  0.044 &
  0.009 &
  0.000 \\
 &
  Topic P &
  \textbf{3.00} &
  \textbf{29.31} &
  \textbf{653.1} &
  \textbf{72.65} &
  \textbf{1.405} \\
 &
  (B\_Topic) &
  (auxiliary &
  (problems) &
  (program &
  (problems) &
  (auxiliary \\
 &
   &
  functions) &
  \multicolumn{1}{l}{} &
  design) &
  \multicolumn{1}{l}{} &
  functions) \\ \hline
\multirow{4}{*}{\begin{tabular}[c]{@{}c@{}}WEEK 2\\ (IN-\\ CLASS)\end{tabular}} &
  R. Time &
  1.013 &
  0.690 &
  0.174 &
  0.002 &
  0.019 \\
 &
  Topic P &
  \textbf{6.391} &
  \textbf{8.488} &
  \textbf{568.8} &
  \textbf{1.779} &
  \textbf{1.592} \\
 &
  (B\_Topic) &
  (problems) &
  (boolean &
  (problems) &
  (program) &
  (program) \\
 &
   &
   &
  value) &
   &
  \multicolumn{1}{l}{} &
   \\ \hline
\multirow{4}{*}{\begin{tabular}[c]{@{}c@{}}WEEK 2\\ (OUT-\\ CLASS)\end{tabular}} &
  R. Time &
  0.050 &
  0.086 &
  0.263 &
  0.091 &
  0.325 \\
 &
  Topic P &
  \textbf{5.580} &
  \textbf{29.97} &
  \textbf{241.1} &
  \textbf{2.845} &
  \textbf{17.92} \\
 &
  (B\_Topic) &
  (design &
  (cond &
  (data &
  (body &
  (exercise \\
 &
   &
  method) &
  expression) &
  analysis) &
  expression) &
  problems) \\ \hline
\multirow{4}{*}{\begin{tabular}[c]{@{}c@{}}WEEK 3\\ (IN-\\ CLASS)\end{tabular}} &
  R. Time &
  0.050 &
  0.460 &
  0.114 &
  0.014 &
  0.337 \\
 &
  Topic P &
  \textbf{11.82} &
  \textbf{8.263} &
  \textbf{7.998} &
  \textbf{15.06} &
  \textbf{5.031} \\
 &
  (B\_Topic) &
  (exercise &
  (synthetic &
  (synthetic &
  (sorting) &
  (examples) \\
 &
   &
  problems) &
  data) &
  data) &
   &
   \\ \hline
\multirow{4}{*}{\begin{tabular}[c]{@{}c@{}}WEEK 3\\ (OUT-\\ CLASS)\end{tabular}} & R. Time & 0.023 & 0.001 & 0.213 & 1.428 & 0.1951 \\
 &
  Topic P &
  \textbf{15.17} &
  \textbf{168.3} &
  \textbf{286.8} &
  \textbf{42.27} &
  \textbf{43.13} \\
 &
  (B\_Topic) &
  (templates) &
  (element &
  (structure &
  (exercise &
  (exercise \\
 &
   &
  \multicolumn{1}{l}{} &
  count) &
  element) &
  problems) &
  problems) \\ \hline
\multirow{4}{*}{\begin{tabular}[c]{@{}c@{}}WEEK 4\\ (IN-\\ CLASS)\end{tabular}} &
  R. Time &
  0.816 &
  0.066 &
  0.156 &
  0.755 &
  0.6471 \\
 &
  Topic P &
  \textbf{2.453} &
  \textbf{9.775} &
  \textbf{2181.5} &
  \textbf{34305} &
  \textbf{7.995} \\
 &
  (B\_Topic) &
  (element &
  (problems) &
  (list) &
  (comparison &
  (abstraction) \\
 &
   &
  search) &
   &
   &
  operation) &
   \\ \hline
\multirow{3}{*}{\begin{tabular}[c]{@{}c@{}}WEEK 4\\ (OUT-\\ CLASS)\end{tabular}} &
  R. Time &
  0.0195 &
  \multicolumn{4}{c}{} \\
 &
  Topic P &
  \textbf{2.069} &
  \multicolumn{4}{c}{Insufficient data} \\
 &
  (B\_Topic) &
  (abstraction) &
  \multicolumn{4}{c}{} \\ \hline
\end{tabular}
\end{table}

In all cases considered, we identified a topic (B\_Topic) that achieved a higher FDR than the Reading Time feature. This result indicates that the differences between two groups of students with different grades are better explained by their preference for specific topics than by their overall engagement with the lecture material. It also suggests that ML models can benefit from using information about topic preferences to predict at-risk students.

The presented results suggest that integrating LECTOR into traditional models enhances the understanding of students’ behaviors by providing new, contextualized information. This new information can be used in two main ways. First, it can be integrated into ML models to improve both predictive performance and explainability of results. Second, it can be used in traditional analyses of reading behaviors \cite{akcapinar_chen_2020,akcapinar_hasnine_2020} to provide richer context for interpreting student engagement patterns.

\subsection{At-risk Students' Prediction Formulation}\label{sec7_3}
In this evaluation, we measured the performance of a simple ML model (Logistic Regression) in predicting at-risk students from two different data representations (independent variable). The first was a vector of activity frequencies F (traditional representation) and the second a vector of topic preferences T (LECTOR integration). Traditional representations considered features such as reading time, page navigation, bookmark, highlight, and note-taking activities (Table \ref{table_tadd2}) \cite{chen_2021,akcapinar_2019}, while the topic preferences considered the relative reading time (Figure \ref{fig:fig_11}) on the most important topics in the lecture (calculated from the $M_ij$ score).

Since a vector of topic preferences corresponds to one lecture, each course provided about 14 datasets (each collected from one of the seven lectures, in-class, and out-class, courses A-2019 to A-2022; Table \ref{table_dataset}) for predicting at-risk students (60 cases in total). For each dataset, we evaluated the model performances (F-score and AUC) in a 3-fold cross-validation with 20 samples for each fold (evaluation proposed by \cite{chen_2021}).

We compared the performance of the different representations and validated the difference with a T-test. For example, the results in Table \ref{table_2_1} correspond to the predictions in the first week of the course A-2019. These last results show that out of class, topic preferences (T) achieved a significantly higher AUC than traditional representations (F), while in the other cases there was no significant difference ($p<0.05$). 

\begin{table}[h]
\centering
\caption{Original results from the first week of the course A-2019.}
\label{table_2_1}
\begin{tabular}{cccll}
\hline
\textbf{Metric} & \textbf{Period}               & \textbf{T}                               & \multicolumn{1}{c}{\textbf{F}} & \textbf{p-value} \\ \hline
AUC     & IN-CLASS & 0.610±0.101 & 0.576±0.158 & 0.155 \\
                & \multicolumn{1}{l}{OUT-CLASS} & \multicolumn{1}{l}{\textbf{0.738±0.099}} & \textbf{0.613±0.178}           & \textbf{6.21E-6} \\ \hline
F-score & IN-CLASS & 0.588±0.088 & 0.597±0.125 & 0.648 \\
                & \multicolumn{1}{l}{OUT-CLASS} & \multicolumn{1}{l}{0.642±0.113}          & 0.603±0.119                    & 0.067            \\ \hline
\end{tabular}
\end{table}

\begin{table}[h]
\centering
\caption{Comparison of prediction performances for different student' representations.}
\label{table_5}
\begin{tabular}{lllccc}
\hline
                                 &                      &         & $\boldsymbol{T>F}$ & $\boldsymbol{F>T}$ & $\boldsymbol{T=F}$ \\ \hline
\multirow{4}{*}{\textbf{A-2019}} & \multirow{2}{*}{IN}  & AUC     & 4           & 0           & 4            \\
                                 &                      & F-score & 0           & 0           & 8            \\ 
                                 & \multirow{2}{*}{OUT} & AUC     & 6           & 0           & 2            \\
                                 &                      & F-score & 3           & 0           & 5            \\ \hline
\multirow{4}{*}{\textbf{A-2020}} & \multirow{2}{*}{IN}  & AUC     & 4           & 1           & 2            \\
                                 &                      & F-score & 2           & 1           & 4            \\
                                 & \multirow{2}{*}{OUT} & AUC     & 2           & 0           & 5            \\
                                 &                      & F-score & 2           & 0           & 5            \\ \hline
\multirow{4}{*}{\textbf{A-2021}} & \multirow{2}{*}{IN}  & AUC     & 1           & 2           & 5            \\
                                 &                      & F-score & 5           & 1           & 2            \\
                                 & \multirow{2}{*}{OUT} & AUC     & 8           & 0           & 0            \\
                                 &                      & F-score & 1           & 0           & 7            \\ \hline
\multirow{4}{*}{\textbf{A-2022}} & \multirow{2}{*}{IN}  & AUC     & 5           & 0           & 2            \\
                                 &                      & F-score & 4           & 0           & 3            \\
                                 & \multirow{2}{*}{OUT} & AUC     & 5           & 0           & 2            \\
                                 &                      & F-score & 4           & 0           & 3            \\ \hline
\end{tabular}
\end{table}

\begin{figure}[h]
  \centering
  \includegraphics[height=5.8in, width=4.2in]{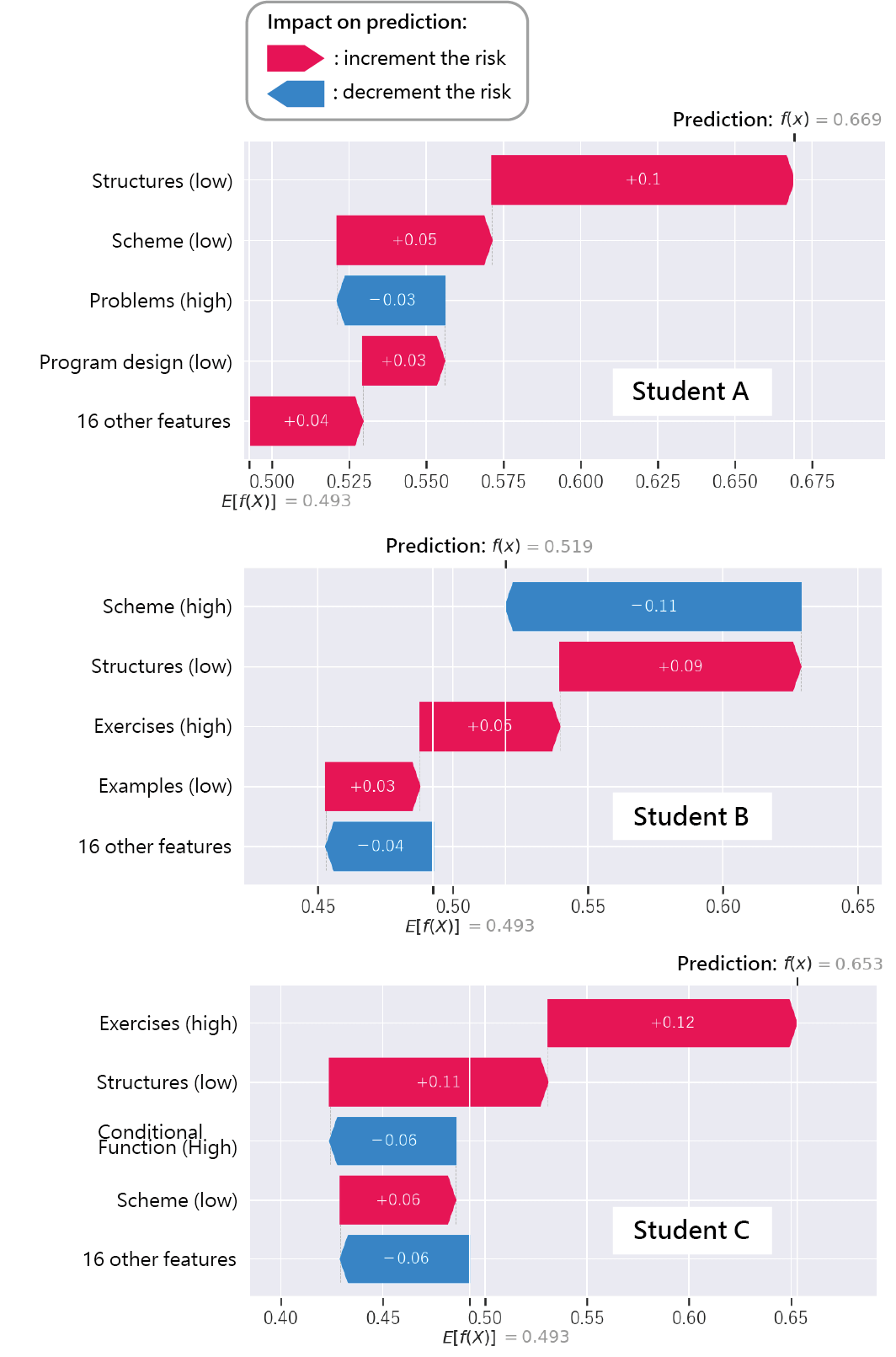}
  \caption{Contribution of topic-based features for three at-risk students in course A-2019.}
  \label{fig:fig_14}
\end{figure}

\subsection{At-risk Students' Prediction Results}\label{sec7_4}
The results from the 60 different cases are summarized in Table \ref{table_5}. Statistical significance was determined using a T-test ($p<0.05$). In this table, the columns represent:
\begin{itemize}
    \item $T>F$: The number of cases where Topic preferences (T) significantly outperformed traditional representations (F). 
    \item $F>T$: The number of cases where traditional representations (F) significantly outperformed Topic preferences (T).
    \item $T=F$: The number of cases where the T-test did not reject the null hypothesis in the previous cases.
\end{itemize}

Because each dataset provided independent evidence of model performance, we did not apply a p-value correction. Accordingly, our results do not constitute a meta-analysis of our independent experiments, but rather a robust aggregation of their evidence.

The low values in column $F>T$ indicate that considering information about topic preferences often leads to an equal or better performance than traditional representations. In particular, in the predictions from OUT-CLASS data, no instances of performance degradation are found, suggesting that topic preferences are more informative when students are not instructed on which part of the material to read.

The values in columns $T>F$ and $T=F$ show that in almost half of the cases the topic preferences information improves the predictions. However, even in the neutral cases ($T=F$), the consideration of topic preferences is beneficial. By following the methodology proposed by \cite{murata_2023} for analyzing the importance of features, we can provide personalized feedback to possible at-risk students. For example, Figure \ref{fig:fig_14} shows the key features (topic preferences) that led to three different students (A, B, and C) being identified as at-risk students. 

A straightforward interpretation suggests that Student A should focus on reading more about \textit{“Structures”} and \textit{“Program Design”}, while also addressing the \textit{“problems”} associated to these topics. Student B should also concentrate on \textit{“Structures”}, but with an emphasis on reviewing \textit{“examples”} rather than \textit{“exercises”}. Student C, on the other hand, should avoid spending excessive time on \textit{“exercises”} and instead revisit the fundamentals of \textit{“Structures”} and \textit{“Conditional Function”}. These insights highlight how the explainability of ML models trained on LECTOR's topic-based features can support instructors in designing personalized educational interventions tailored to individual student needs.

\section{Discussion}\label{sec8}
\subsection{Experiment 1}\label{sec8_1}
Our two evaluations validate the effectiveness of incorporating educational considerations –specifically, the intrinsic hierarchical structure of slides -- into an information retrieval system for extracting key information from lecture slides. LECTOR outperformed state-of-the-art NLP models and traditional educational tools in keyphrase extraction, achieving an average F1-score of 20.14\% compared to AttentionRank \cite{ ding_luo_2021} (15.06\%) and TF-IDF \cite{ salton_1988} (15.86\%), a widely used backbone in educational tools.

Traditional unsupervised keyphrase extraction models, such as TF-IDF, TextRank, and AttentionRank \cite{salton_1988, mihalcea_2004,ding_luo_2021}, have reported F1-scores ranging from 10\% to 40\% on benchmark datasets written in narrative-style text, such as Inspec, SemEval2017, and DUC2001. However, our dataset consists of lecture slides, which lack a conventional narrative structure and often contain semantic noise, such as fragmented text, bullet points, and flowcharts. These characteristics negatively impact the performance of general NLP models, especially AttentionRank \cite{ding_luo_2021} (Table \ref{table_tadd3}).

The degradation of AttentionRank’s performance can be attributed to two key limitations in its design:

\begin{description}
\item[\textbf{1. Influence of Word Frequency:}] AttentionRank’s Accumulated Self-Attention module prioritizes frequently occurring words. While beneficial for large documents, this becomes a drawback in lecture slides, where domain-specific terms appear frequently in examples. For instance, terms such as “define” and “else” are commonly used in programming exercises but do not represent key instructional concepts.
\item[\textbf{2. Contextualization in Cross-Attention:}] AttentionRank’s Cross-Attention module assigns higher scores to keyphrases with explicit contextual references in slide content. This leads to a bias toward selecting keyphrases directly referenced in text snippets, rather than conceptually important terms. For example, in a slide discussing list processing, AttentionRank may favor explicit terms from a code snippet (e.g., “define”) while failing to extract core instructional concepts such as “list processing” or “example code”.
\end{description}

In contrast, LECTOR overcomes these limitations through its specialized design:

\begin{description}
\item[\textbf{1. Reducing the Influence of Word Frequency:}] LECTOR incorporates the SIF (Smooth Inverse Frequency) factor \cite{arora_2017}, which down-weights frequent words and instead prioritizes structurally significant topics that align with educational discourse.
\item[\textbf{2. Addressing Contextualization Bias:}] Unlike AttentionRank, LECTOR leverages the intrinsic hierarchical structure of slides within its cross-attention mechanism, reformulated as a probability estimation model. This ensures that concepts central to the lecture’s discourse receive higher relevance scores, leading to more accurate and educationally meaningful keyphrase extractions.
\end{description}

While these benefits are primarily technical, they translate into significant educational advantages, particularly for the development of new educational tools. The practical impact of LECTOR is demonstrated by the human evaluation results, which indicate that LECTOR’s selected keyphrases align more closely with students' perceptions of relevant information on each lecture slide compared to traditional methods. This suggests that LECTOR is not only effective in computational keyphrase extraction but also in identifying concepts that students find meaningful for learning.

This practical improvement in information retrieval has important implications for educational applications, including:

\begin{description}
\item[\textbf{1. Slide Summarization Systems:}] LECTOR’s ability to extract core instructional concepts that align with student expectations can enhance automated summarization tools, helping students efficiently review key lecture points \cite{shimada_2018}. Additionally, LECTOR’s estimation of the content covered in each slide enables the generation of customized summaries based on a specified list of topics.
\item[\textbf{2. Recommendation Systems:}] By aligning more closely with students’ perceptions, LECTOR can enhance the quality of material recommendations. Additionally, by integrating topic-based reading preferences, LECTOR improves the contextualized information used in content recommendation engines, resulting in better-personalized recommendations \cite{nakayama_2019, okubo_2020}.
\end{description}

Although LECTOR demonstrates superior performance compared to previous information retrieval systems, there are still certain limitations to its effectiveness in this task:
\begin{description}
\item[\textbf{1. Dependence on Well-Structured Slides:}] While leveraging the hierarchical discourse of slides is one of LECTOR's primary strengths, it also introduces a limitation. LECTOR may extract inaccurate relationships from lecture slides with unclear titles or poorly defined hierarchies. Consequently, LECTOR's performance is closely tied to the instructor’s ability to define clear slide hierarchies and descriptive titles.
\item[\textbf{2. Limited Contextual Information:}] LECTOR’s current use of hierarchical information does not incorporate contextual information from preceding or subsequent slides, which is often critical for comprehending the content of the current slide. This limitation arises from the challenges of dynamically identifying slides that discuss the same topic.
\item[\textbf{3. Text-Only Focus:}] LECTOR relies exclusively on textual information, making it less effective for slides that predominantly feature images, diagrams with minimal text, mathematical notations, and other non-textual elements.
\end{description}

To expand LECTOR’s capabilities, future work should aim to enhance its ability to contextualize information and integrate multimodal content. Improving contextualization will require extensive analysis of mechanisms that effectively capture the continuity between slides. Meanwhile, addressing the text-only limitation could involve employing recent advancements in vision-language models, such as GPT-4 Omni and Gemini \cite{openAI,gemini}, to interpret non-textual content and incorporate this data into the textual analysis.

\subsection{Experiment 2}\label{sec8_2}
The results of our two evaluations validate that integrating LECTOR into traditional models to contextualize reading activities provides new insights into student behavior, improving both the performance and explainability of at-risk student predictive models. In 100\% of the cases, groups of students with different performance levels focused on different topics despite reading for a similar amount of time. This characteristic allows LECTOR to improve ML model performance (average F1-score increase of 4.67\%) while also revealing patterns relevant for educational intervention (e.g., SHAP values of topic-based features).

This improvement in predictive performance aligns with prior research on using contextualized NLP models to identify at-risk or dropout students \cite{cadd_1,cadd_2,cadd_4,cadd_6,cadd_7,cadd_8}. Despite methodological differences, all these studies found improvements in predicting at-risk indicators, such as confusion levels \cite{cadd_1}, bug-fix time \cite{cadd_2}, cognitive presence \cite{cadd_6}, and direct dropout likelihood \cite{cadd_4, cadd_7, cadd_8}. Similar to our findings, these improvements were attributed to the rich contextual information that textual data provides when integrated into structured features.

A common practice in these studies is using NLP pre-trained models as feature extractors, where embeddings from BERT \cite{cadd_1, cadd_6, cadd_7}, CodeBERT \cite{cadd_2}, and XLNet or TF-IDF \cite{cadd_8} were employed to classify at-risk students. While our study also leverages a pre-trained model (BERT), our approach differs in that LECTOR follows an information retrieval methodology, estimating how much information about different topics each lecture slide contains rather than directly encoding textual meaning. This methodology enhances explainability, as it generates interpretable features that allow educators to understand which concepts influence the model’s classification of at-risk students, unlike prior studies that rely on dense vector embeddings, which lack direct semantic interpretability.

Using pre-trained models as feature extractors provides a more direct approach to at-risk student classification and potentially yields to higher raw predictive performance. However, our proposed additional step of information retrieval has significant implications for personalized educational interventions. For instance:

\begin{description}
\item[\textbf{1.}] LECTOR can be integrated into at-risk student classifiers to identify students struggling with specific key concepts.
\item[\textbf{2.}] The extracted concepts (and their SHAP importance values) can be used to generate a summarized document that proportionally focuses on those concepts, leveraging LECTOR for slide summarization (see Section \ref{sec8_1}).
\item[\textbf{3.}] The same concepts can be used in recommender systems implementing LECTOR to suggest additional resources tailored to student needs (see Section \ref{sec8_1}).
\end{description}

Accordingly, our proposed approach helps to close the loop in at-risk student prediction, ensuring that interventions are targeted and actionable. In addition, the use of LECTOR topic-based features can enable researchers to conduct more context-rich analyses of students’ reading behaviors. Previous studies have used clustering analyses to identify general behaviors that align with student learning frameworks \cite{akcapinar_chen_2020,akcapinar_hasnine_2020,yin_2019}. Future works can extend this approach by integrating topic-based features, following a similar methodology to our Basic Separability Analysis (see Section \ref{sec7_1}).

For instance, a previous study \cite{akcapinar_chen_2020} applied clustering techniques within the Student Approaches to Learning (SAL) framework \cite{marton_76}, identifying three student groups that followed Deep, Surface and Strategic approaches. The Deep Approach was associated with the most engaged students, the Strategic Approach with students active during classes but not outside classes, and the Surface Approach with students who exhibited low engagement during classes and no engagement outside. However, a key limitation of this study is that a Surface Approach involves selective engagement with content necessary for upcoming assessments, requiring additional semantic information to fully distinguish this cluster.

In this context, our Basic Separability Analysis (Section \ref{sec7_2}) provides insights that complement the findings described above. Our results on Table \ref{table_4} show that, during weeks two and three, at-risk students exhibit reduced engagement outside of class, a pattern consistent with the identified Surface Approach \cite{akcapinar_chen_2020}. However, our findings further reveal that these students focused primarily on "exercise problems" (B\_Topic), validating the idea that this group of students prioritize content directly related to evaluations rather than engaging with broader conceptual material. This highlights the value of integrating topic-based features into student behavior analyses to gain deeper insights into learning patterns.

While we highlighted the benefits of LECTOR topic-based features for educational interventions and learning analytics, our current approach has some limitations:
\begin{description}
\item[\textbf{a. Higher dimensionality of features:}] LECTOR extracts fewer features than embedding-based approaches \cite{cadd_1, cadd_2, cadd_6, cadd_7, cadd_8}, however, it still increases feature dimensionality compared to traditional features. For instance, a single feature such as reading time is expanded into multiple topic-specific features (e.g., relative reading time for 50 different topics). Since a higher number of features increases the risk of overfitting, larger datasets are required for more robustness.
\item[\textbf{b. Dependence on Information Retrieval Quality:}] The keyphrase estimation step introduces a potential source of noise, as errors in information retrieval can propagate into topic-based features. For example, an implementation of TF-IDF in this step may include frequent but irrelevant words as topic features. Therefore, this module needs to be optimized independently before it is integrated into the reading logs.
\item[\textbf{c. Lack of Temporal Modeling:}] Previous research \cite{cadd_2} has demonstrated that tracking changes in textual representations (CodeBERT embeddings) over time improves the identification of struggling students. Similarly, studies using traditional reading log features \cite{lopez_icce,murata_2023} have shown that time-aware models improve predictive performance by capturing sequential learning behaviors. However, our current implementation only analyzes single-week snapshots of student activity, limiting its ability to track how students’ engagement with different topics evolves. This limitation arises because the topics covered change from week to week, making it difficult to develop a consistent representation of student learning trajectories.
\item[\textbf{d. Exclusive Use of Topic-Based Features:}] Unlike previous models that concatenate textual and structured features \cite{cadd_2, cadd_7, cadd_8}, our implementation only integrates textual information into new features without explicitly incorporating traditional engagement metrics. While topic-based features capture content focus, they do not directly measure engagement, which remains a critical predictor of student performance.
\end{description}

Accordingly, future research should address these limitations by improving feature engineering and refining the estimation of slide-topic scores. While the refinement of the retrieval system has already been discussed in Section \ref{sec8_1}, further improvements are needed in the definition and integration of topic-based features. This includes redefining topic-based features, integrating them with engagement features, and adapting them to time-aware models. The redefinition of topic-based should focus on reducing dimensionality and ensuring standardization across time. One possible approach is to implement a hierarchical knowledge graph, allowing for adjustable topic granularity and enabling interconnections between topics from different weeks. Meanwhile, the concatenation with engagement metrics and integration into time-aware models primarily require technical development and do not present significant methodological challenges.

\subsection{Limitations}
In addition to the limitations of LECTOR discussed in the previous section (Sections \ref{sec8_1}, \ref{sec8_2}), the present study should be considered in light of the following limitations:

\begin{description}
\item[\textbf{1. Limited domain of lecture slides:}] Both experiments were conducted using lecture slides from the "Programming Theory" course, delivered in Japanese, restricting the generalizability of our findings. To assess the broader applicability of LECTOR, future studies should evaluate its performance on slides from different courses and languages.

\item[\textbf{2. Indirect Evaluation of the Model Estimates:}] In Experiment 1 (technical evaluation, Section \ref{sec6_2}), an ideal evaluation should assess retrieval capabilities at the slide-topic level. However, due to the impracticality of collecting such labels (Section \ref{sec6_1}), we instead conducted an indirect evaluation using keyphrase extraction. While the second part of the experiment (human evaluation, Section \ref{sec6_3}) further supports LECTOR's superiority, this evaluation was also limited by the number of participants (28 students) and its associated cost (USD 2,000).

\item[\textbf{3. Limited Analysis of Topic-based Features:}] In Experiment 2 (Basic separability analysis, section \ref{sec7_2}), since results across years are not aggregable, we restricted our analysis to data from 2019 to avoid an overwhelming number of results. While this aspect of Experiment 2 is limited in scope, its practical implications for at-risk student prediction were validated across multiple years in the second evaluation (prediction performance, section \ref{sec7_4}) where we aggregated results to assess performance improvements.
\end{description}

\section{Conclusions}\label{sec9}
We proposed LECTOR, a new model that adapts state-of-the-art keyphrase extraction techniques to the domain of lecture slides. Thanks to its internal re-contextualization of the information, LECTOR is more effective at extracting key information from lecture slides than the predominant model TF-IDF and the state-of-the-art model AttentionRank. The information extracted by LECTOR is valuable for improving existing educational tools, such as slide summarization systems, recommendation systems, and e-book footprints' preservation.

The topic preferences extracted by LECTOR encapsulate information that can partially explain differences in students' final grades. Incorporating this new information into traditional ML models for predicting at-risk students often results in equal or better performance. Furthermore, by analyzing the contribution of the different preferences to the prediction results, these new models can help identify representative reading patterns of at-risk students (e.g., surface learning approaches) and inform the design of personalized interventions to help students comprehend the reading materials.

Considering the wide range of potential applications of models like LECTOR, future research could consider addressing the design of more powerful retrieval methods to extract information from lecture slides. 

Specifically, the current design of LECTOR poses two main technical challenges. The first is improving the contextualization of the information in the slides by considering contextual windows of slides and correcting inaccurate titles. The second is extracting more nuanced and actionable topics by updating topic extraction methods. To approach these challenges, a promising way is to integrate the advances in current Large Language Models (LLM) in future developments.

From a more practical perspective, further research could explore the use of “content-based” reading characteristics in analyses of reading behavior and at-risk prediction to design educational feedback for students and teachers. We suggest the following possibilities: 
\begin{itemize}
    \item Correcting surface learning approaches and recommending supplemental learning materials based on reading preferences.
    \item Guidance for reviewing specific reading content to improve the expected grades.
    \item Support student cognition by showing connections between key concepts and slides.
    \item Support student metacognition by displaying personal and class reading topic tracks.
    \item Help teachers refine current learning materials based on reading logs.
    \item Identify in real-time what content students are struggling with in the lecture.
\end{itemize}

\bmhead{Acknowledgements}
This work was supported by JST CREST Grant Number JPMJCR22D1 and JSPS KAKENHI Grant Number JP22H00551, Japan.

\section*{Declarations}
\subsection*{Funding}
This work was supported by JST CREST Grant Number JPMJCR22D1 and JSPS KAKENHI Grant Number JP22H00551, Japan.

\subsection*{Competing interests}
The authors have no competing interests to declare that are relevant to the content of this article.

\subsection*{Author Contribution Information}

E.L. contributed to the conceptualization of the study, data curation, formal analysis, methodology development, investigation, and visualization. He also wrote the original draft and participated in the review and editing process. C.T. and V.Š. were responsible for supervision and contributed to reviewing and editing the manuscript. F.O. provided supervision, managed the project, and contributed to the review and editing of the manuscript. A.S. oversaw the supervision and project administration, contributed to the funding acquisition, validation, and methodology, and participated in reviewing and editing the manuscript.

\subsection*{Data availability}
The datasets including reading logs and lecture materials are not available (Table \ref{table_dataset}) as our consent forms did not include information regarding sharing data outside of the research study. Upon a reasonable direct request, we can provide specially anonymized versions of them. The complete data generated in the four experiments included in this article can be consulted in the following GitHub: \href{https://github.com/limu-research/LECTOR_V1}{experiment results}.

\bibliography{sn-bibliography}

\end{document}